\documentclass[usenatbib, useAMS]{mnras}
\usepackage{natbib}
\usepackage{graphicx}
\usepackage{rotating}
\usepackage{verbatim}
\usepackage{amssymb}
\usepackage{amsmath}
\usepackage{gensymb}
\usepackage{pdflscape}
\usepackage{dcolumn}
\usepackage[toc,page]{appendix}
\newcolumntype{d}[1]{D{.}{.}{#1}}

\begin{document}

\title[The Baryonic Tully-Fisher relation]
{From light to baryonic mass: the effect of the stellar mass--to--light ratio
on the Baryonic Tully--Fisher relation}
\author[Ponomareva et al.]
{Anastasia A. Ponomareva$^{1,2}$\thanks{E-mail:
anastasia.ponomareva@anu.edu.au}, Marc A. W. Verheijen$^{2,3}$, Emmanouil Papastergis$^{2,5}$,\newauthor
 Albert Bosma$^4$ and Reynier F. Peletier$^2$ \vspace*{0.2cm}\\
   $^1$Research School of Astronomy \& Astrophysics, Australian National University, Canberra, ACT 2611,
  Australia\\
  $^2$Kapteyn Astronomical Institute, University of Groningen, Postbus 800, 
  NL-9700 AV Groningen, The Netherlands\\
  $^3$National Centre for Radio Astrophysics, Tata Insttitute of Fundamental Research, Postbag 3, Ganeshkhind, Pune 411 007, India \\
  $^4$Aix Marseille Univ, CNRS, LAM, Laboratoire d'Astrophysique de Marseille, Marseille, France\\
  $^5$Credit Risk Modeling Department, Co\"{o}perative Rabobank U.A., Croeselaan 18, Utrecht NL-3521CB, The Netherlands \label{rabo}\\}

\pagerange{\pageref{firstpage}--\pageref{lastpage}}
\pubyear{2017}

\maketitle
\label{firstpage}
\begin{abstract}

  In this paper we investigate the statistical properties of the
  Baryonic Tully-Fisher relation (BTFr) for a sample of 32 galaxies
  with accurate distances based on Cephe\"ids and/or TRGB stars.  We
  make use of homogeneously analysed photometry in 18 bands ranging
  from the FUV to 160 $\mu$m, allowing us to investigate the effect of
  the inferred stellar mass--to--light ratio ($\Upsilon_{\star}$) on
  the statistical properties of the BTFr. Stellar masses of our sample
  galaxies are derived with four different methods based on full
  SED--fitting, studies of stellar dynamics ,
  near-infrared colours, and the
  assumption of the same $\Upsilon_{\star}^{[3.6]}$ for all galaxies.  In
  addition, we use high--quality, resolved H{\sc i} kinematics to
  study the BTFr based on three kinematic measures: $W^{i}_{50}$ from
  the global HI profile, and $V_{max}$ and $V_{flat}$ from the
  rotation curve.  We find the intrinsic perpendicular scatter, or
  tightness, of our BTFr to be $\sigma_{\perp}=0.026\pm 0.013$ dex,
  consistent with the intrinsic tightness of the 3.6 $\mu$m
  luminosity--based TFr. However, we find the slope of the BTFr to be
  2.99$\pm$0.2 instead of 3.7$\pm$0.1 for the luminosity--based TFr at
  3.6$\mu$m.  We use our BTFr to place important observational
  constraints on theoretical models of galaxy formation and evolution
  by making comparisons with theoretical predictions based on either the 
  $\Lambda$CDM  framework or modified Newtonian dynamics.

\end{abstract}


\defcitealias{TC12}{TC12}
\defcitealias{pon16}{P16}
\defcitealias{pon17}{P17}

\section{Introduction}

The empirical scaling relations of galaxies are a clear demonstration
of the underlying physical processes that govern the formation and evolution of
galaxies. Any particular theory of galaxy formation and evolution
should therefore explain their origin and intrinsic properties such as
their slope, scatter and zero point. One of the most multifunctional
and well--studied empirical scaling relations is the relation between the width
of the neutral hydrogen line and the luminosity of a galaxy
\citep{tf77}, known as the Tully-Fisher relation (TFr).  Originally
established as a tool to measure distances to galaxies, it became one
of the most widely used relations to constrain theories of galaxy
formation and evolution \citep{ns00, vog14, schaye15, verbeke15,
  nihao16}.  Even though the TFr has been extensively studied and
explored during the past four decades \citep{v01, mcgaugh05, TC12,
sor13, karachentsev17}, many open questions still remain, especially those
relating to the physical origin and the underlying physical mechanisms
which maintain the TFr as galaxies evolve \citep{mcgaugh98,
  courteau99, vdbosch00}.  Finding answers to these questions is
crucial for our comprehension of galaxies and how they form and
evolve.

At present, the physical principle behind the TFr is widely considered
to be a relation between the baryonic mass of a galaxy and the mass of
its host dark matter (DM) halo \citep{milgrom88, freeman99, mcgaugh05}
since the TFr links the baryonic content of a galaxy (characterised by
its luminosity) to a dynamical property (characterised by its
rotational velocity).  Therefore, if a galaxy's luminosity is a proxy
for a certain baryonic mass fraction, a relation between its
rotational velocity and its total baryonic mass should exist.  Indeed,
\citet{mcgaugh00} have shown the presence of such a relation, which is
now widely known as the Baryonic Tully-Fisher relation (BTFr).

Subsequently, the BTFr was widely studied \citep{bell01, zaritsky14,
  manolistfr, lelli16} as it has a great potential to put quantitative
constraints on models of galaxy formation and evolution.  Moreover, it
clearly offers some challenges to the $\Lambda$CDM cosmology
model. Foremost, it follows just a single power--law over a broad
range of galaxy masses. This is contrary to the expected relation in
the $\Lambda$CDM paradigm of galaxy formation, where the BTFr
``curves" at the low velocity range \citep{TG11,desmond12}. Second,
the BTFr appears to be extremely tight, suggesting a zero intrinsic
scatter \citep{v01, mcgaugh12}. It is difficult to explain such a
small observational scatter in the BTFr, as various theoretical
prescriptions in simulations, such as the mass--concentration relation
of dark matter halos or the baryon--to--halo mass ratio, result in a
significant scatter. For instance, \citet{lelli16} have found an
intrinsic scatter of $\sim$ 0.1 dex, while \citet{dutton12} predicts a
minimum intrinsic scatter of $\sim$ 0.15 dex, using a semi-analytic
galaxy formation model in the $\Lambda$CDM context. However,
\citet{manolistfr} have shown that theoretical results seem to
reproduce the observed BTFr better if hydrodynamic simulations are
considered instead of semi-analytical models
\citep{governato12,brooks14,christ14}.  This suggests that the
galactic properties that are expected to contribute to the intrinsic
scatter (halo spin, halo concentration, baryon fraction) are not
completely independent from each other.  Moreover, the BTFr is also
used to test alternative theories of gravity, and various studies
argue that the observed properties of the BTFr can be better explained
by a modification of the gravity law (e.g. MOND, \citealp{milgrom83})
than by a theory in which the dynamical mass of galaxies is dominated
by the DM, such as in the $\Lambda$CDM scenario.

The BTFr can be considered as a reliable tool to test galaxy formation
and evolution models only if the statistical properties of the BTFr
compared between observations and simulations are as consistent as
possible.  So far, various observational results differ in details,
even though they find similar results in general. For instance, the
slope of the observed relation varies from $\sim$ 3.5
\citep{bell01,zaritsky14} to $\sim$ 4.0
\citep{mcgaugh00,manolistfr,lelli16}.  Therefore, it is important to
address the observational limitations when studying the BTFr because
the measurements of the rotational velocity and of the baryonic mass
of galaxies are rather difficult.  The baryonic mass of a galaxy is
usually measured as the sum of the stellar and neutral gas components.
According to \citet{bland16}, the contribution of the hot halo gas is
larger in mass, but this is usually not accounted for in the BTFr.  Since
the neutral atomic gas mass can be measured straightforwardly from
21-cm line observations while the contribution of molecular gas to the
total baryonic mass is often small, the biggest contributor to the
uncertainty in the BTFr is the stellar mass measurement. Even though
various prescriptions to determine the stellar mass are available, the
relative uncertainty in the stellar mass derived from photometric
imaging usually ranges between 60-100 \% \citep{pforr12}.  Moreover,
various recipes for deriving the stellar mass-to-light ratio
($\Upsilon_{\star}$) from spectro--photometric measurements depend on a
number of parameters, such as the adopted initial stellar mass
function (IMF), the star formation history (SFH) and uncertainties in
modelling the advanced phases of stellar evolution, such as AGB stars
\citep{maraston06,conroy09}.  There are alternative ways to measure
the stellar mass of galaxies, for example by estimating it from the
vertical velocity dispersion of stars in nearly face--on disk galaxies
\citep{bershady10}.  However, such methods are observationally
expensive and have systematic limitations as well \citep{aniyan16}.

The BTFr requires an accurate measurement of the rotational velocity of
galaxies. There are several methods to estimate this parameter,
derived either from the width of the global H{\sc i} profile or from
spatially resolved H{\sc i} kinematics. It was shown by \citet{v01}
that the scatter in the luminosity--based TFr can be decreased if the
velocity of the outer flat part ($V_{flat}$) of the rotation curve is
used as a measure of the rotational velocity, instead of the corrected
width of the global H{\sc i} profile $W_{50}^{i}$.  As was shown in
\citealp{pon16} (hereafter \citetalias{pon16}), the rotational
velocity derived from the width of the global H{\sc i} profile may
differ from the value measured from the flat part of the rotation
curve, especially for galaxies which have either rising or declining
rotation curves. These issues should be considered when studying the
statistical properties of the BTFr. Likewise, \citet{brook16}
demonstrated with a set of simulated galaxies that the statistical
properties of the BTFr vary significantly, depending on the rotational
velocity measure used.

In order to take into account the uncertainties mentioned above and to
establish a more definitive study of the BTFr, we consider in detail
four methods to estimate the stellar mass of galaxies. This allows us
to study the dependence of the statistical properties of the BTFr as a
function of the method used to determine the stellar mass: full
SED--fitting, dynamical $\Upsilon_{\star}$ calibration,
$\Upsilon_{\star}^{[3.6]}$ as a function of [3.6]-[4.5] colour and
constant $\Upsilon_{\star}^{[3.6]}$. Furthermore, we consider the BTFr
based on three velocity measures: $W_{50}$ from the corrected width of
the global H{\sc i} profile, and $V_{flat}$ and $V_{max}$ from the
rotation curve.  This allows us to study how the slope, scatter and
tightness of the BTFr change if the relation is based on a different
definition of the rotational velocity.

This paper is organised as follows: Section \ref{sample} describes the
sample of BTFr galaxies. Section \ref{data} describes the data
sources. Section \ref{gassection} describes the gas mass
derivation. Section \ref{stmass} describes four methods to estimate
the stellar mass of the sample galaxies. Section \ref{btfrscomp}
provides the comparison of the BTFrs based on different stellar mass
measurements. Section \ref{adoptedbtfr} presents our adopted
BTFr. Section \ref{comp} demonstrates the comparison of our adopted
BTFr with previous observational studies and theoretical
results. Section \ref{concl} presents concluding remarks.

\section{The sample}\label{sample}

In order to study the statistical properties of the BTFr and to be
able to compare our results with the luminosity--based TFr studied in
\citealp{pon17} (hereafter \citetalias{pon17}), we adopt the same
sample of 32 galaxies used in \citetalias{pon17}. These galaxies were
selected according to the following criteria: 1) $Sa$ or later in
morphological type, 2) an inclination above 45$^{\circ}$, 3) H{\sc i}
profiles with adequate S/N and without obvious distortions or
contributions from possible companions to the flux.  The main
properties of the sample are summarised in Table 1 in
\citetalias{pon16}. This sample has been specifically selected to
study the BTFr in detail and has properties that help minimize many of
the observational uncertainties involved in the measurement of the
relation such as: 1) poorly known distances, 2) the conversion of light into
stellar mass, and 3) the lack of high-quality H{\sc i} rotation
curves.

First, galaxies in our sample have accurate primary distance
measurements, either from the Cephe\"id period--luminosity relation
\citep{freedman01} or/and from the brightness of the tip of the red
giant branch \citep{rizzi07}. If simple Hubble flow distances were
used for the nearby galaxies in our sample, the distance uncertainties
might contribute up to 0.4 mag to the observed scatter of the
luminosity--based TFr. In contrast, the distance uncertainty
contribution to the observed scatter in the TFr is only 0.07 mag if
independently measured distances are adopted \citepalias{pon17}.
Second, our adopted sample benefits from homogeneously analysed
photometry in 18 bands, ranging from $FUV$ to 160 $\mu$m. This allows
us to perform a spectral energy distribution (SED) fitting to derive
the stellar masses based on stellar population modelling.  Third, all
galaxies from our sample have H{\sc i} synthesis imaging data and
high--quality rotation curves available, from which we derive the
maximum rotational velocity $V_{max}$ and the outer flat rotational
velocity $V_{flat}$ \citepalias{pon16}.

\section{Data sources}\label{data}

To derive the main ingredients for the BTFr such as stellar masses,
molecular and atomic gas masses and rotational velocities, we use the
following data sources and techniques.

\subsection{21--cm aperture synthesis imaging}

For our study we collected 21--cm aperture synthesis imaging data from
the literature, since many of our galaxies were already observed as
part of several large H{\sc i} surveys (see \citetalias{pon16} for an
overview). Moreover, we observed ourselves three more galaxies with
the GMRT (see \citetalias{pon16} for more details).  All data cubes
were analysed in the same manner and various data products were
derived, including global H{\sc i} profiles, surface density profiles
and high--quality rotation curves.  Based on these homogeneous H{\sc
  i} data products, we have measured rotational velocities in three
ways: from the corrected width of the global H{\sc i} profile
($V_{circ}=W_{50}^{R,t,i}/2$), as the maximal rotational velocity
($V_{max}$) from the rotation curve, and as the velocity of the outer
``flat" part of the rotation curve ($V_{flat}$), noting that massive
and compact galaxies often show a declining rotation curve in their
inner regions where $V_{max} > V_{flat}$.

\subsection {Photometry} \label{phot}

In \citetalias{pon17} we have derived the main photometric properties
of our sample galaxies to study the wavelength dependence of the
slope, scatter and tightness of the luminosity--based TFr in 12
photometric bands from $FUV$ to 4.5 $\mu$m with magnitudes corrected
for internal and Galactic extinction.  For this work, we have
collected and analysed supplementary Wide-field Infrared Survey
Explorer (WISE, \citealp{wise}) imaging data at 12 $\mu$m and 22
$\mu$m. In each passband, we calculate magnitudes in apertures of
increasing area and extrapolate the resulting surface brightness
profiles to obtain the total magnitudes. Moreover, to determine the
far-infrared emission we collected from the literature far-infrared
fluxes at 60 $\mu$m and 100 $\mu$m as measured by IRAS, and at 70
$\mu$m and 160 $\mu$m as measured with Herschel/MIPS. We use these
photometric measurements to estimate the stellar masses of our sample
galaxies.

\section{Gas mass} \label{gassection}

Gas is an important contributor to the baryonic mass of a spiral
galaxy and plays a crucial role in the study of the BTFr. For
instance, when assuming the same stellar mass--to--light ratio for all
sample galaxies, only the gas mass is responsible for any difference
in the slope and tightness of the BTFr compared to the
luminosity--based TFr.

The H{\sc i} mass can be directly measured from the 21-cm radio
observations, while the H$_{2}$ mass can only be obtained indirectly
using either $CO$ or wartm dust observations
\citep{leroy09,westfall11,martinsson13}. Although generally the atomic
gas mass dominates over the molecular component, there are several
known cases where the estimated mass of the molecular gas is similar
to, or exceeds, the mass of the atomic gas
\citep{leroy09,saintonge11,martinsson13}.  Therefore, it is important
to take both constituents into account when studying the BTFr.  In
this section we describe how the masses of the atomic and molecular
gas were derived.

\subsection{H{\sc i} mass}

We calculate the H{\sc i} masses of our sample galaxies using the
integrated H{\sc i}--line flux density ($\int S_{\nu} dv$ [Jy
kms$^{-1}$]) derived as part of the analysis of the 21-cm radio
synthesis imaging observations \citepalias{pon16}, according to:

\begin{equation}
M_{HI}\,[M_{\odot}]= 2.36 \times 10^{5} \times D^2[Mpc]\, \int S_{\nu} dv\, [Jy kms^{-1}],
\end{equation}

\noindent
where $D$ is the distance to the galaxy, as listed in
\citetalias{pon16}, Table 1.  We derive the error on the H{\sc i} mass
by following a full error propagation calculation, taking into account
the measurement error on the flux density and the error on the
distance modulus as calculated in \citetalias{pon16} .  Furthermore,
we calculate the total neutral atomic gas mass as

\begin{equation}
M_{atom}= 1.4 \times M_{HI},
\end{equation}

\noindent
where the factor 1.4 accounts for the primordial abundance of helium
and metals.  The mass of the neutral atomic gas component is listed in
Table \ref{tbl_masses}.  It is important to note that we estimate the
H{\sc i} mass under the assumption that all of the 21 cm emission is
optically thin, which is not always the case and up to 30\% of the H{\sc
  i} mass can be hidden due to H{\sc i} self--absorption according
to \citet{peters17}.

\subsection{H$_{2}$ mass}

\begin{figure}
\begin{center}
\includegraphics[scale=0.70]{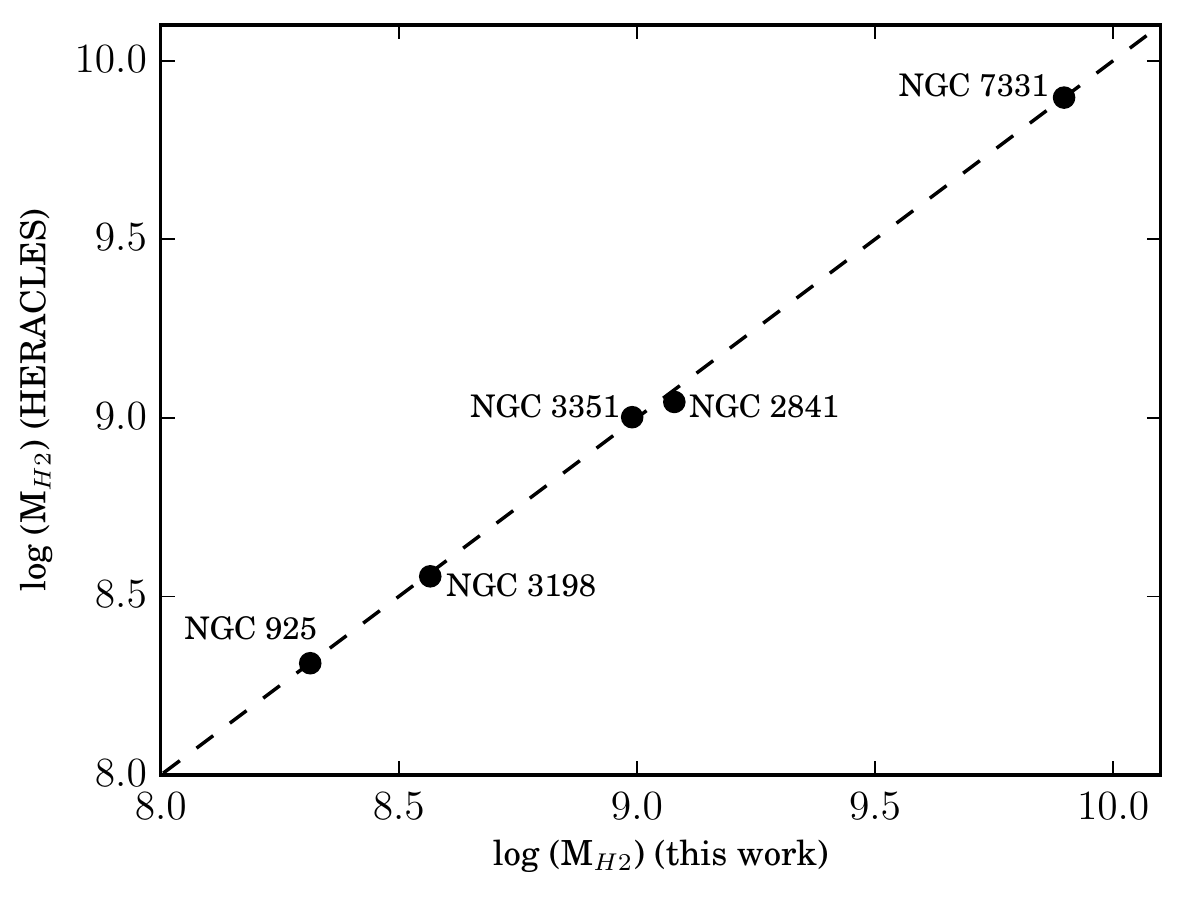}
\caption{The comparison between the H$_{2}$ mass derived using the 22
  $\mu$m surface brightness (this work), and the H$_{2}$ mass derived
  from direct $CO$ measurements from the HERACLES survey
  \citep{leroy09}.  The dashed line represents the 1:1
  correspondence.}
\label{fig_mh2}
\end{center}
\end{figure}

Unfortunately, the distribution of the molecular hydrogen (H$_{2}$) in
galaxies can not be directly observed. Therefore, indirect methods are
required to estimate the mass of the H$_{2}$ (M$_{H_{2}}$). The most
straightforward and widely studied tracer of the H$_{2}$ gas is the
$CO$ emission line, which can be directly observed
\citep{leroy09,saintonge11, young91}.  The M$_{H_{2}}$ can be
estimated using the $^{12}CO (J=1 \rightarrow 0)$ flux
($I_{CO} \Delta V$) and the $^{12}CO (J=1 \rightarrow 0)$-to-H$_{2}$
conversion factor $X_{CO}$. However, only 5 out of 32 galaxies in our
sample have $CO$ measurements available.  In order to ensure a
homogeneous analysis, we use instead the 22 $\mu$m imaging photometry
to estimate the $CO$ column--density distribution.  This approach is
motivated by various studies that demonstrate a tight correlation
between the infrared luminosity of spiral galaxies, associated with
the thermal dust emission, and their molecular gas content as traced
by the $CO$ emission \citep{westfall11, bendo07, paladino06, young91}.
For our study we use the following relation from \citet{westfall11} to
derive $I_{CO} \Delta V$:
\begin{figure}
\begin{center}
\includegraphics[scale=0.60]{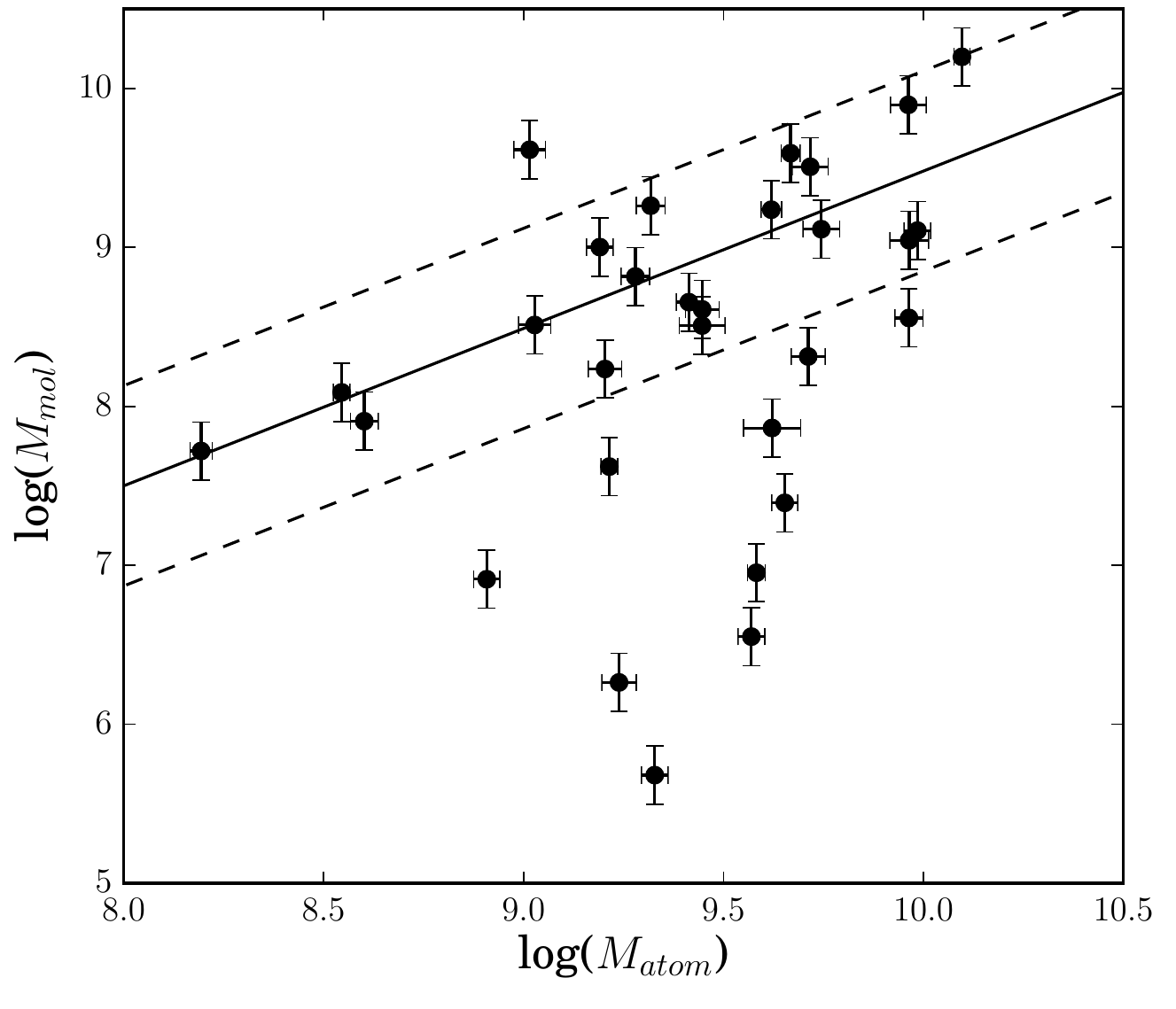}
\caption{$M_{atom}$ vs. $M_{mol}$ for our sample galaxies. The solid
  line indicates the fit from \citet{saintonge11}. The dashed lines
  represent the scatter in the $M_{atom}-M_{mol}$ relation
  ($\sigma=0.41$ dex) also from \citet{saintonge11}. Note that only 30
  galaxies are shown, as we did not detect NGC 3319 and NGC 4244 at
  22$\mu$m.}
\label{fig_gasmas}
\end{center}
\end{figure}

\begin{equation}
log (I_{CO} \Delta V) = 1.08 \cdot log(I_{22\mu m})+0.15,
\end{equation}

\noindent
where $I_{CO} \Delta V$ is in K kms$^{-1}$ and $I_{22\mu m}$ is the 22
$\mu$m surface brightness in MJy sr$^{-1}$. Note that
\citet{westfall11} used 24 $\mu$m fluxes in their study. However, the
24 $\mu$m and 22 $\mu$m bands are very similar and therefore we
proceed our study using the 22 $\mu$m flux. We do not detect 22 $\mu$m flux
emission in only 2 galaxies, NGC 3319 and NGC 4244, which is not surprising
as these two galaxies are dwarfs. In dwarf galaxies, the low metallicities
cause large uncertainties in the $X_{CO}$ factor, which makes it impossible to relate the amount of $CO$ 
to $H_{2}$. Of course, H$_{2}$ is strongly correlated with the SFR, which is very low in dwarf galaxies, and therefore 
one can conclude that the contribution of $H_{2}$ to the bulk mass of a dwarf galaxy is negligible.
\begin{figure*}
\begin{center}
\includegraphics[scale=0.55]{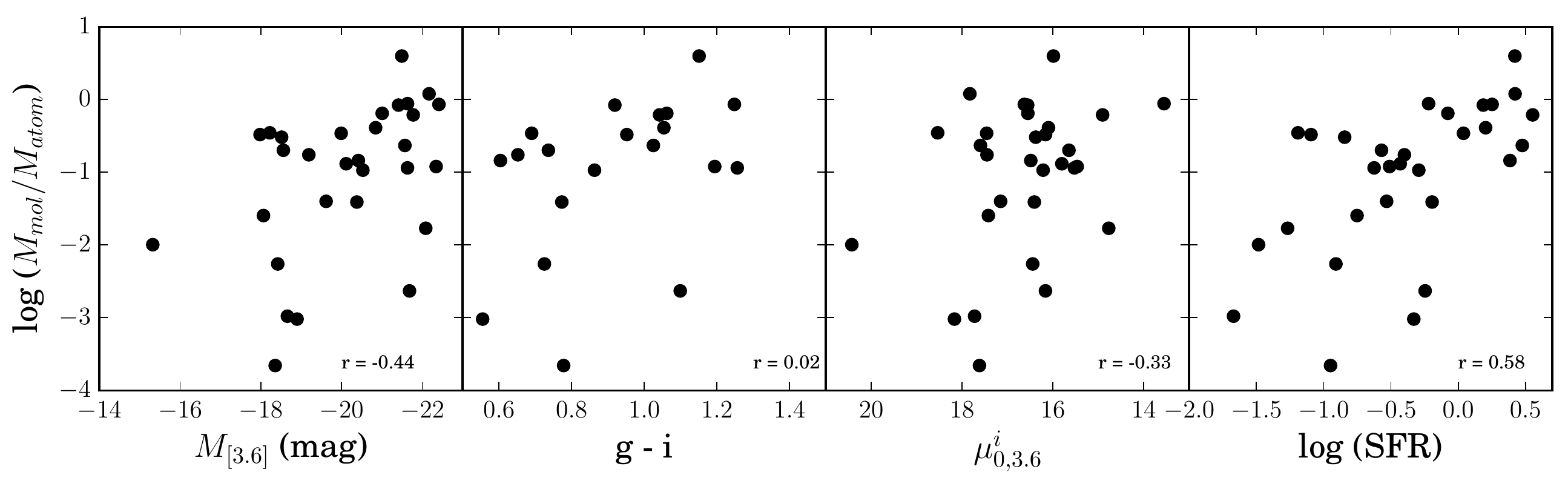}
\caption{Correlations between $R_{mol}$ and global galaxy
  properties. In the bottom right corner the Pearson's correlation
  coefficients are shown.}
\label{fig_gascor}
\end{center}
\end{figure*}
Next, we calculate the M$_{H_{2}}$ using the $X_{CO}$ conversion
factor \citep{westfall11,martinsson13}:

\begin{equation}
 \Sigma M_{H_{2}} [M_{\odot}pc^{-2}] = 1.6 I_{CO} \Delta V \times X_{CO} \cdot cos(i) ,
\end{equation}
\noindent
where $i$ is the inclination angle as derived from the H{\sc i}
kinematics.  Even though the use of $X_{CO}$ is a standard procedure
to convert $CO$ column density into molecular hydrogen gas mass,
different studies offer various derivations of $X_{CO}$. Here, we
adopt $X_{CO}=2.7(\pm0.9) \times 10^{20} cm^{-2} (K kms^{-1})^{-1}$
from \citet{westfall11}. In that study, they use a mean value of the
Galactic measurement of $X_{CO}$ from \citet{dame01} and the
measurements for M31 and M33 from \citet{bolatto08}.  In Figure
\ref{fig_mh2} we compare our H$_{2}$ masses with those derived from
the direct $CO$ measurements from the HERACLES survey \citep{leroy09}
for five galaxies in our sample.  It is clear that our estimates are
in good agreement.  To account for the mass of helium and heavier
elements corresponding to hydrogen in the molecular phase, we
calculate the mass of the molecular gas component as:

\begin{equation}
M_{mol}= 1.4 \times M_{H_{2}}.
\end{equation}

\noindent
Despite a good agreement with the HERACLES measurements, the method to
estimate the $CO$ column density from the 22 $\mu$m surface brightness
has its limitations which result in a significant estimated error on
the molecular gas mass of $\sim$ 42 \%
\citep{westfall11,martinsson13}, which we adopted for our
measurements.

\subsection{ M$_{atom}$ vs. M$_{mol}$}

Presuming that the molecular gas forms out of collapsing clouds of
atomic gas, it seems reasonable to expect a tight correlation between
the masses of the atomic and molecular components. However, recent
studies of the gas content of large galaxy samples have shown that
this is not the case. A large scatter is present in the
$M_{atom}$--$M_{mol}$ relation
\citep{leroy09,saintonge11,martinsson13}. Figure \ref{fig_gasmas}
shows the $M_{atom}$--$M_{mol}$ relation for our sample galaxies. Even
though the majority of our galaxies follows the relation from
\citet{saintonge11} with a similar scatter, we have some outliers with
smaller molecular--to--atomic mass ratio ($R_{mol}=M_{mol}/M_{atom}$)
which lie below the bottom dashed line in Figure \ref{fig_gasmas}.

In Figure \ref{fig_gascor} we present correlations between $R_{mol}$
and global galaxy properties such as absolute magnitude, colour,
central surface brightness and star formation rate. Even though there
are some hints that more luminous, redder galaxies with a higher star
formation rate tend to have a larger fraction of $M_{mol}$, the
scatter in these correlations is very large, with the best correlation
between $logR_{mol}$ vs SFR. In general, $R_{mol}$ for individual
galaxies ranges greatly from 0.001 to 3.97 with a mean value of
$<R_{mol}>=0.38$, which is in good agreement with previous studies
\citep{leroy09,saintonge11,martinsson13}.  We find one extreme case,
NGC 3627, with $R_{mol}=3.97$, comparable to UGC 463 with
$R_{mol}=2.98$ \citep{martinsson13}, NGC 4736 with $R_{mol}=1.13$
\citep{leroy09} and G38462 with $R_{mol}=4.09$ \citep{saintonge11}.

It is important to mention that in this section we deliberately do not
compare masses of the gaseous components with the estimated masses of
the stars in our sample galaxies, because the measurement of the
stellar masses is not straightforward and the contribution of the
stellar mass to the baryonic mass budget can vary, depending on the
method used to estimate stellar masses. We discuss this subject in the
following section.

\section{Stellar masses} \label{stmass}
The stellar masses of galaxies, unlike the light, can not be measured
directly and, therefore, their estimation is a
very complex process with various assumptions and uncertainties. The
most common method of estimating the stellar mass of a galaxy is to
convert the measured light into mass using a relevant mass--to--light
ratio. However, deciding which mass--to--light ratio to use is not
straightforward. It can be derived either from stellar population
synthesis models or by measuring the dynamical mass (surface) density
of a galaxy.  Every method of estimating the mass-to-light ratio has its 
uncertainties and limitations. In this paper we consider four different methods 
of estimating the stellar masses, and work out for each of them their effect on 
the statistical properties of the BTFr. In our study, regardless of the method,
we limit ourselves to integral mass--to--light ratios, 
i.e. no radial variation of this quantity is analysed. Even though bulges are expected to have
higher mass--to--light ratios than disks, we justify this approach given
the absence of strong colour gradients with radius and the small number of 
bulge dominated galaxies in our sample.
\begin{figure*}
\begin{center}
\includegraphics[scale=0.50]{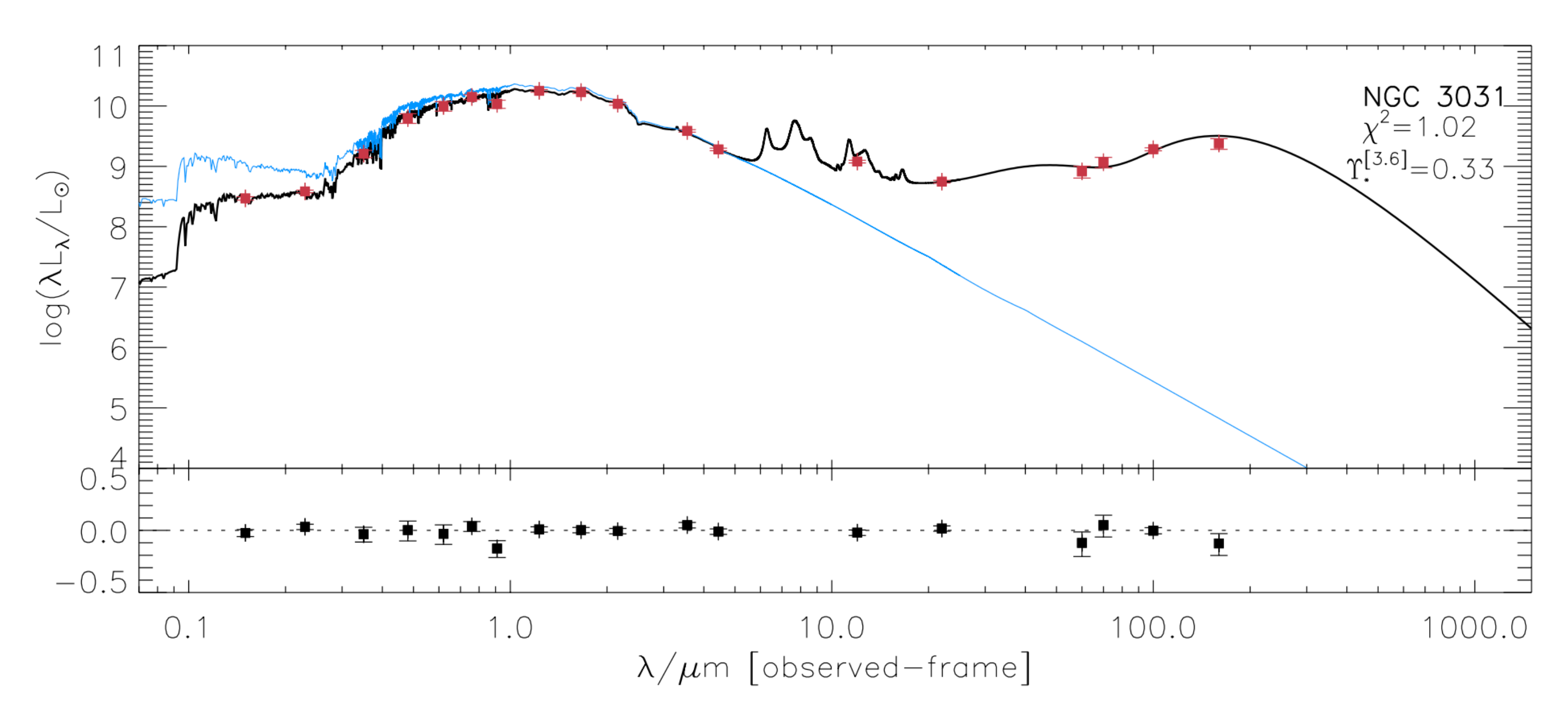}
\caption{ An example of the best--fit model, performed with MAGPHYS
  (in black) over the observed spectral energy distribution of NGC
  3031. The blue curve shows the unattenuated stellar population
  spectrum.  The bottom plot shows the residuals for each measurement
  ($(L_{obs}-L_{mod})/L_{obs}$).}
\label{fig_sed}
\end{center}
\end{figure*}
\subsection{SED modeling}
The light that comes from stars of different ages and masses dominates
the flux in different photometric bands. Thus, for example, young hot
stars dominate the flux in the $UV$ bands while old stellar
populations are more dominant in the infrared bands: e.g. 0.8-5
$\mu$m. Moreover, mid-- and far--infrared bands can trace the galactic
dust at various temperatures.  The differences between magnitudes in
these photometric bands (galactic colours) contain information on
various properties of the stars in a galaxy such as their age or
metallicity.  Therefore, stellar population models aim to create a mix
of stellar populations that is able to simultaneously reproduce a wide
range of observed colours.  Hence, modelling of the spectral energy
distribution allows us to estimate the total stellar mass of the
composite stellar population.  This process is called spectral energy
distribution (SED) fitting.

It is important to measure the luminosity of a galaxy at as many
wavelengths as possible in order to provide more constraints on the
various physical parameters of a model. Having photometric
measurements in many bands, spanning from the far--ultraviolet to the
far--infrared, helps to derive more reliable values for various
galactic properties that influence the estimation of the stellar mass
(e.g., star formation history, metallicity). However, it should be
kept in mind that stellar mass estimates from SED--fitting are
nonetheless limited by systematic uncertainties in the theoretical
modelling of stellar populations. For example, limited knowledge
regarding the initial mass function (IMF), uncertainties in the
theoretical modelling of advanced stages of stellar evolution, or
limitations of stellar spectral libraries cannot be suppressed with
better photometric data.

To calculate the stellar masses of our sample galaxies using
SED--fitting, we derived fluxes in 14 photometric bands from $FUV$ to
22 $\mu$m.  Moreover, we collected from the literature far--infrared
fluxes at 60 $\mu$m and 100 $\mu$m as measured by IRAS, and at 70
$\mu$m and 160 $\mu$m, as measured with Herschel/MIPS (see Section
\ref{phot}). Consequently, we have at our disposal measured fluxes in
18 photometric bands for every galaxy (except for 10 galaxies that
lack SDSS data, see \citetalias{pon17}).  We performed the fitting of
the spectral energy distribution of every galaxy, using the
SED--fitting code ``MAGPHYS", following the approach described in
\citet{dacunha08}.  The advantage of this code is its ability to
interpret the mid-- and far--infrared luminosities of galaxies
consistently with the UV, optical and near--infrared luminosities.  To
interpret stellar evolution it uses the \citet{bruzual03} stellar population synthesis model. 
This model predicts the spectral evolution
of stellar populations at ages between $1 \times 10^{5}$ and
$2 \times 10^{10}$ yr. In this model, the stellar populations of a
galaxy are described with a series of instantaneous bursts, so called
``simple stellar populations". The code adopts the \citet{chabrier03}
Galactic disk IMF. The code also takes into account a new prescription
for the evolution of low and intermediate mass stars on the thermally
pulsating asymptotic giant branch \citep{marigo07}. This prescription
helps to improve the prediction of the near-infrared colours of an
intermediate age stellar population, which is important in the context
of spiral galaxies.  To describe the attenuation of the stellar light
by the dust, the code uses the two--component model of
\citet{charlot00}. It calculates the emission from the dust in giant
molecular clouds and in the diffuse ISM, and then distributes the
luminosity over wavelengths to compute the infrared spectral energy
distribution.  The ability of the SED--fitting code to take a dusty
component into account while performing the stellar mass estimate is
very important for our study because we deal with star forming spirals
in which the amount of dust and obscuration is not negligible.
From the SED fitting we derive a stellar mass estimate for each galaxy
in our sample. As an example, the best--fit SED model for NGC 3031 is
shown in Figure \ref{fig_sed} (see Appendix \ref{appendix:sed} for the
SED fits of other sample galaxies).

\begin{table*}
\begin{tabular}{lrrrccc}
\hline
name     &  log(sSFR)& log(SFR)& log(M$_{\star}^{SED})$ &log(M$_{dust})$& $\Upsilon_{\star}^{SED,[3.6]}$ & $\Upsilon_{\star}^{SED, K}$\\
\hline
NGC 55	    & -11.60& -2.40 &9.20  & 6.27 & 0.32&0.34\\
NGC 224	    & -15.37& -4.79 &10.57 & 7.16 & 0.21&---\\
NGC 247	    & -10.55& -1.51 &9.04  & 6.68 & 0.15&0.21\\
NGC 253	    & -13.06& -2.64 &10.41 & 8.03 & 0.22&0.32\\
NGC 300	    & -10.57& -1.66 &8.91  & 6.58 & 0.19&0.21\\
NGC 925	    & -12.18& -2.38 &9.80  & 7.08 & 0.35&0.41\\
NGC 1365	& -13.18& -2.38 &10.80 & 7.56 & 0.33&0.38\\
NGC 2366	& -7.81 & -0.65 &7.16  & 4.57 & 0.04&0.04\\
NGC 2403	& -10.77& -1.61 &9.16  & 6.86 & 0.12&0.17\\
NGC 2541	& -10.89& -1.90 &8.99  & 6.09 & 0.16&0.23\\
NGC 2841	& -14.36& -3.69 &10.67 & 8.05 & 0.21&0.27\\
NGC 2976	& -10.79& -1.94 &8.84  & 6.07 & 0.17&0.24\\
NGC 3031	& -13.97& -3.38 &10.59 & 7.70 & 0.33&0.45\\
NGC 3109	& -9.27 & -1.62 &7.65  & 4.71 & 0.14&0.19\\
NGC 3198	& -11.83& -1.94 &9.89  & 7.56 & 0.21&0.27\\
IC 2574      	& -8.30 & -0.58 &7.72  & 5.63 & 0.03&0.06\\
NGC 3319	& -11.33& -1.99 &9.34  & 6.28 & 0.24&0.26\\
NGC 3351	& -12.75& -2.40 &10.35 & 7.34 & 0.34&0.40\\
NGC 3370	& -10.73& -1.17 &9.56  & 7.25 & 0.10&0.14\\
NGC 3621	& -12.18& -2.25 &9.93  & 7.47 & 0.30&0.40\\
NGC 3627	& -11.99& -1.79 &10.21 & 7.93 & 0.16&0.20\\
NGC 4244	& -10.62& -1.79 &8.83  & 5.95 & 0.12&0.18\\
NGC 4258	& -12.77& -2.40 &10.37 & 7.24 & 0.21&0.26\\
NGC 4414	& -12.52& -1.98 &10.54 & 8.21 & 0.26&0.31\\
NGC 4535	& -12.50& -2.13 &10.37 & 7.74 & 0.25&0.30\\
NGC 4536	& -12.25& -2.06 &10.19 & 7.86 & 0.28&0.41\\
NGC 4605	& -11.07& -1.82 &9.25  & 6.64 & 0.26&0.36\\
NGC 4639	& -12.51& -2.36 &10.15 & 7.18 & 0.34&0.41\\
NGC 4725	& -13.57& -2.91 &10.66 & 7.66 & 0.38&0.41\\
NGC 5584	& -10.93& -1.46 &9.47  & 6.54 & 0.12&0.12\\
NGC 7331	& -13.93& -2.84 &11.09 & 8.20 & 0.52&0.68\\
NGC 7793	& -11.64& -2.27 &9.37  & 6.38 & 0.36&0.44\\

\hline
\end{tabular}
\caption{Results of the SED-fitting performed with MAGPHYS.
Column (1): name;
Column (2): log of the specific star formation rate;
Column (3): log of the star formation rate;
Column (4): log of the stellar mass;
Column (5): log of the dust mass;
Column (6): stellar mass-to-light ratio for the stellar masses from Column (4) and light in the 3.6$\mu$m band;
Column (7): stellar mass-to-light ratio for the stellar masses from Column (4) and light in the K-- band;
}
\label{tbl_sedfits}
\end{table*} 

\begin{figure}
\begin{center}
\includegraphics[scale=0.55]{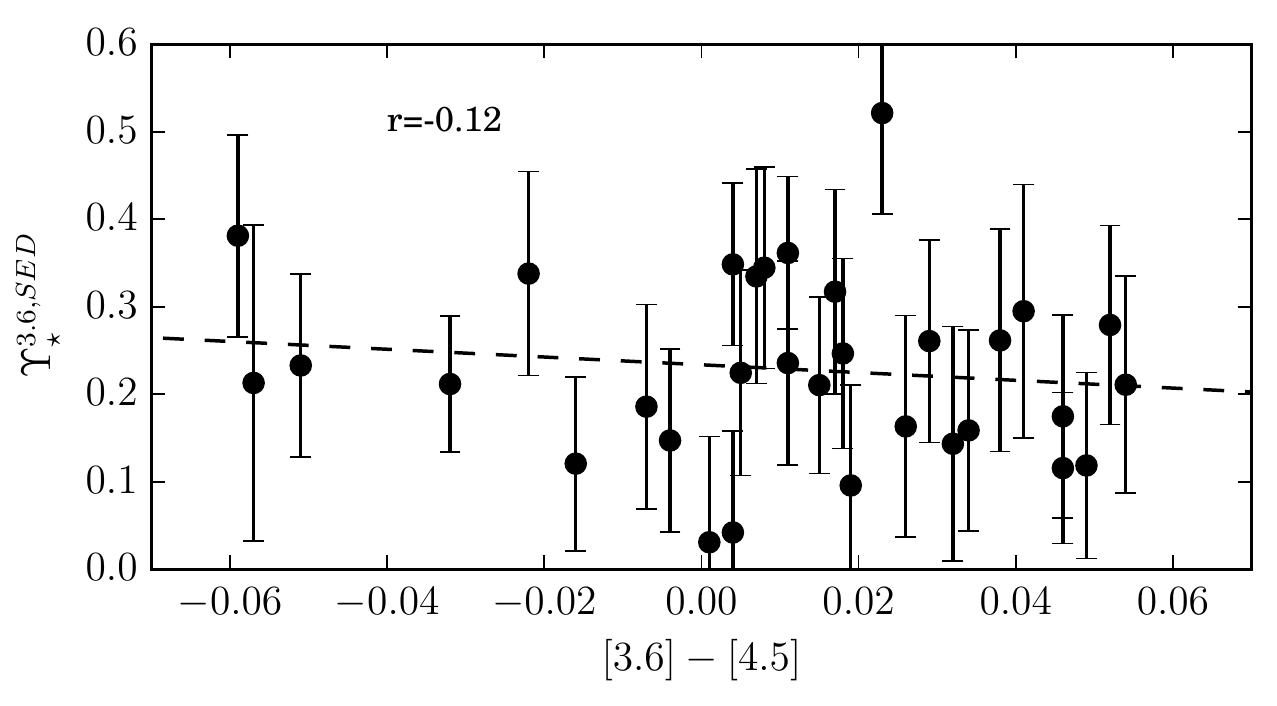}
\caption{ Derived stellar mass-to-light ratios from the SED--fitting
  ($\Upsilon_{\star}^{SED,[3.6]}$) as a function of the $[3.6]-[4.5]$
  colour. The linear fit is shown with the dashed line, $r$ is Pearson's correlation coefficient.}
\label{fig_mlSEDcol}
\end{center}
\end{figure}

Thereby we obtain the stellar mass--to--light ratio
($\Upsilon_{\star}$) for the light in several photometric bands. 
 We present the $\Upsilon_{\star}$ for the $K$ and 3.6 $\mu$m bands in Table
\ref{tbl_sedfits} together with the other parameters obtained from the
SED modelling.  Notably, we will refer to the stellar mass--to--light
ratio, measured from the SED-fitting as
$\Upsilon_{\star}^{SED, \lambda}$, where $\lambda$ is a particular
photometric band. Interestingly, we do not find any correlation
between $\Upsilon_{\star}^{SED,[3.6]}$ and the $[3.6]-[4.5]$ colour
(Figure \ref{fig_mlSEDcol}), while such a correlation exists in case
$\Upsilon_{\star}^{[3.6]}$ is measured with other methods (see below).

We assign a relative error to the SED--based stellar mass--to--light
ratio ($\Upsilon_{\star}^{SED,\lambda}$) equal to
$\epsilon_{\Upsilon_{\star}^{SED,\lambda}}=0.1$ dex motivated by the
test by \citet{roediger15}, who performed SED--fitting with ``MAGPHYS"
on a sample of mock galaxies. They could recover the known stellar
masses with a scatter of 0.1 dex for various samples using a different
number of observational bands.  Finally, we calculate a fractional
error on the stellar mass as follows:

\begin{equation}
\label{stmasserreq}
\epsilon_{M_{\star}^{SED}}^{2}=(10^{\epsilon_{m}/2.5}-1)^2+\epsilon_{\Upsilon_{\star}}^2 ,
\end{equation}  

\noindent
where $\epsilon_{m}$ is the mean error in the absolute magnitude over
all bands, equal to $\epsilon_{m}=0.15$ mag.  Note that the distance
uncertainty is already included in this error on the magnitude.  The
global parameters of our sample galaxies based on the SED--fitting
method are summarised in Table \ref{tbl_sedfits}.

\begin{figure}
\begin{center}
\includegraphics[scale=0.55]{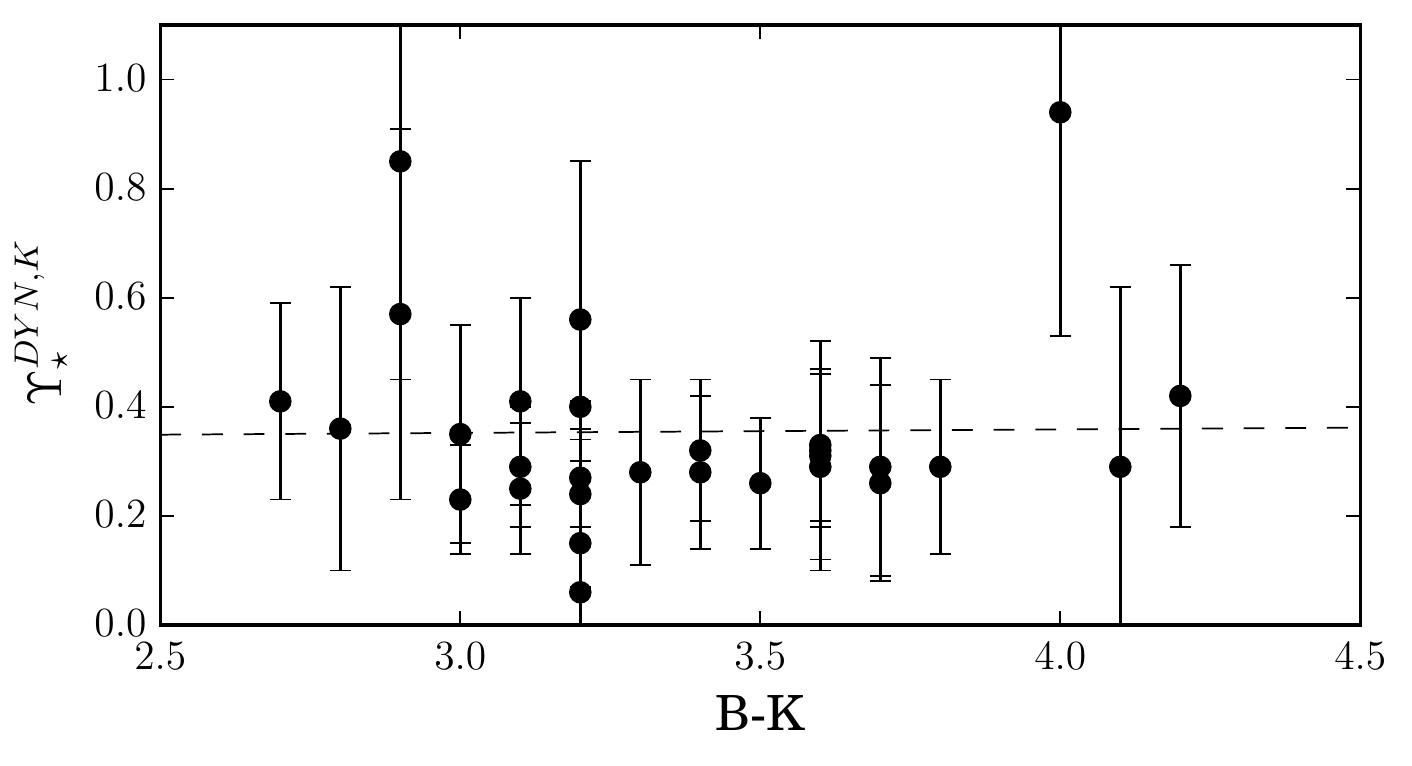}
\caption{ Stellar mass-to-light ratios from the DMS
  ($\Upsilon_{\star}^{Dyn, K}$) as a function of the B-K colour.}
\label{fig_mldm}
\end{center}
\end{figure}

\subsection{Dynamical $\Upsilon_{\star}$ calibration }

Another method to estimate the stellar masses of spiral galaxies is by
measuring the mass surface density of their disks dynamically. The
strategy for disk galaxies is to measure the vertical stellar velocity
dispersion ($\sigma_{z}$), which can be used to obtain the dynamical
mass surface density of a collisionless stellar disk in
equilibrium:

\begin{equation}
\Sigma_{dyn} = \frac{\sigma_{z}^{2}}{\pi G \kappa h_{z} \mu},
\end{equation}

\noindent
where $\mu$ is the surface brightness, G is the gravitational
constant, $h_{z}$ is the disk scale height and $\kappa$ is the
vertical mass distribution parameter \citep{vdkruit81,bahcall84}.
While $\mu$ can be easily measured from photometric studies, and there
is a well--calibrated relation between the disk scale length $h_{r}$
and disk scale height $h_{z}$ \citep{degrijs96,kregel02}, $\sigma_{z}$
is very difficult to measure. Here, we take advantage of the DiskMass
Survey (DMS) \citep{bershady10} for dynamically calibrated stellar
mass-to-light ratios, which were obtained by measuring $\sigma_{z}$
for a sample of 30 spiral galaxies \citep{martinsson13}. The DMS
sample is mostly $Sc$ spirals and does not overlap with our BTFr
sample.  In that study the line-of-sight stellar velocity dispersion
($\sigma_{LOS}$) was measured and then converted into $\sigma_{z}$. To
minimize errors on $\sigma_{z}$, which significantly affect
$\Sigma_{dyn}$, spiral galaxies close to face--on were observed.
Consequently, the stellar mass surface density was calculated
following:

\begin{equation}
\Sigma_{\star} = \Sigma_{dyn}-\Sigma_{mol}-\Sigma_{atom},
\end{equation}

\noindent
where $\Sigma_{mol}$ and $\Sigma_{atom}$ are the mass surface
densities of the molecular and atomic hydrogen, see Section
\ref{gassection}. Then, the stellar mass-to-light ratio
$\Upsilon_{\star}$ can be expressed as follows:

\begin{equation}
\Upsilon_{\star} =\frac{\Sigma_{\star}}{\mu},
\end{equation}

\noindent
where $\mu$ is the K-band surface brightness \citep{martinsson13}.
We refer to this stellar mass-to-light ratio as $\Upsilon_{\star}^{Dyn, K}$.

Next, we use $\Upsilon_{\star}^{Dyn, K}$ from the DMS and check if
those values correlate with a colour term, which can be measured
directly from the photometry.  If such a correlation exists, we would
be able to adopt the $\Upsilon_{\star}^{Dyn, K}$ as a function of
colour for our sample.  However, we did not find any correlation ( see
Figure \ref{fig_mldm}). Therefore, we adopt a median value for
$\Upsilon_{\star}^{Dyn, K}$ from \citet{martinsson13}, equal to
$<\Upsilon_{\star}^{Dyn, K}>=0.29$ and we apply it to our K--band
magnitudes to derive stellar masses for our sample galaxies:

\begin{equation}
M_{\star}^{Dyn} =<\Upsilon_{\star}^{Dyn, K}> \cdot L_{K}(L_{\odot}),
\end{equation}

\noindent
where the absolute luminosity of the Sun in the K--band is equal to
3.27 mag.  For the error on $<\Upsilon_{\star}^{Dyn, K}>$ we adopt the
median error from \citet{martinsson13} equal to
$\epsilon_{<\Upsilon_{\star}^{Dyn, K}>}=0.19$ dex and then we
calculate the fractional error on the stellar mass according to
Eq. \ref{stmasserreq}. We estimate the error on our magnitudes as the
mean error of the K--band apparent magnitude, equal to
$\epsilon_{m}=0.17$ mag.

\subsection{$\Upsilon_{\star}^{[3.6]}$ as a function of [3.6]--[4.5] colour}

The flux in the 3.6 $\mu$m band is considered a good tracer of the old
stellar population of galaxies, which is the main contributor to the
total stellar mass, especially in early--type galaxies (ETGs).
Therefore, in recent years much attention has been given to finding
the best way to convert the 3.6 $\mu$m flux into stellar mass
\citep{eskew12,miguel14,rock15,meidt12}. Many of these studies found a
correlation between $\Upsilon_{\star}^{[3.6]}$ and the $[3.6]-[4.5]$
colour.

For instance, \citet{eskew12} used measurements of the resolved Large
Magellanic Cloud (LMC) star formation history (SFH) \citep{harris09}
to calibrate $\Upsilon_{\star}^{[3.6]}$ by linking the mass in various
regions of the LMC to the 3.6 $\mu$m flux. They found that the stellar
mass can be traced well by the 3.6 $\mu$m flux if a bottom-heavy
initial mass function (IMF), such as Salpeter, or heavier was
assumed. They estimated the stellar mass-to-light ratio to be
$\Upsilon_{\star}^{[3.6]}=0.54$ with a 30 \%
uncertainty. Subsequently, they found that $\Upsilon_{\star}^{[3.6]}$
in each region of the LMC correlates with the local $[3.6]-[4.5]$
colour, according to:

\begin{equation} \label{esqew}
log \Upsilon_{\star}^{[3.6]} = -0.74 ([3.6]-[4.5]) - 0.23. 
\end{equation}

\noindent
Hence, Eq. \ref{esqew} can be applied to calculate the stellar masses
of our galaxies, if the fluxes at 3.6 $\mu$m and 4.5 $\mu$m are known.

However, it was demonstrated by \citet{meidt12} that the flux in the
3.6 $\mu$m band can be contaminated by non--stellar emission from warm
dust and from PAHs \citep{shapiro10}. Therefore, they applied an
Independent Component Analysis (ICA) to separate the 3.6 $\mu$m flux
into contributions from the old stellar population and from
non-stellar sources. Thus, according to \citet{meidt14} and
\citet{norris14}, a single $\Upsilon_{\star}^{[3.6]}= 0.6$ can be used
to convert the 3.6 $\mu$m flux into stellar mass, with an uncertainty
of only $0.1$ dex, provided the observed flux is corrected for
non--stellar contamination.  Remarkably, a constant
$\Upsilon_{\star}^{[3.6]}= 0.6$ was also found by stellar population
synthesis models in the infrared wavelength range (2.5--5 $\mu$m),
using empirical stellar spectra \citep{rock15}. In addition,
\citet{miguel14} presented an empirical calibration of
$\Upsilon_{\star}^{[3.6]}$ as a function of $[3.6]-[4.5]$ colour for
galaxies for which the correction for non--stellar contamination was
applied. Thus, they expressed the corrected stellar mass-to-light
ratio as

\begin{equation}
\Upsilon_{\star}^{[3.6],corr} =(\Upsilon_{\star}^{[3.6]}=0.6) \times \frac{F_{[3.6], cor}}{F_{[3.6], uncor}},
\end{equation}

\noindent
where $F_{[3.6], cor}$ is the total 3.6 $\mu$m flux corrected for
non--stellar contamination and $F_{[3.6], uncor}$ is the observed
total flux. Hence, a constant $\Upsilon_{\star}^{[3.6]}=0.6$ is
applicable to observed galaxies without any non-stellar contamination,
such as in ETGs, while $\Upsilon_{\star}^{[3.6]}$ will decrease for those
galaxies which suffer the most from contamination, such as
star--forming spirals.  Furthermore, they expressed
$\Upsilon_{\star}^{[3.6], cor}$ as a function of the $[3.6]-[4.5]$
colour according to:

\begin{equation} \label{icaml}
log \Upsilon_{\star}^{[3.6], cor}= -0.339(\pm 0.057) ([3.6]-[4.5]) - 0.336(\pm 0.002).
\end{equation}

As shown in \citetalias{pon17}, the scatter in the luminosity--based
TFr can be reduced if the corrected 3.6 $\mu$m luminosities are used.
Therefore, we prefer Eq. \ref{icaml} for the calibration of
$\Upsilon_{\star}^{[3.6]}$ as a function of [3.6]--[4.5] colour. In
the remainder of this text, we refer to this mass--to--light ratio as
$\Upsilon_{\star}^{[3.6], cor}$.  We assign an error to
$\Upsilon_{\star}^{[3.6], cor}$ equal to:

\begin{equation}
\epsilon_{\Upsilon_{\star}^{ [3.6], cor}}^{2}=\epsilon_{\Upsilon_{\star}^{3.6}=0.6}^2+\epsilon_{F_{[3.6], cor}/F_{[3.6], uncor}}^2,
\end{equation}

where $\epsilon_{\Upsilon_{\star}^{3.6}=0.6}$ is equal to 0.1 dex
\citep{meidt14} and $\epsilon_{F_{[3.6], cor}/F_{[3.6], uncor}}$ is an
averaged error on the flux ratios at 3.6 $\mu$m, equal to 0.1 dex.
Furthermore, we calculate the fractional error on the stellar mass
according to Eq. \ref{stmasserreq}, using the error on the magnitude
as the mean error on the 3.6 $\mu$m apparent magnitude, equal to
$\epsilon_{m}=0.08$ mag.
\subsection{Constant $\Upsilon_{\star}^{[3.6]}$}
Despite all previously listed motivations to assign different stellar
mass--to--light ratios to disk galaxies, various studies advocate the
use of a single mass-to-light ratio for the 3.6 $\mu$m flux.
Different stellar population modelling results estimate
$\Upsilon_{\star}^{[3.6]}$ in the range between 0.42 \citep{mcgaugh12,
  schombert14} and 0.6 \citep{rock15,meidt14,norris14}, pointing out
that it is metallicity--dependent.  \citet{mcgaugh16} argue that
assigning a universal $\Upsilon_{\star}^{[3.6]}$ allows for a direct
representation of the data with minimum assumptions, while other
methods introduce many more uncertainties.

Furthermore, \citet{lelli16} studied the statistical properties of the
BTFr with resolved H{\sc i} kinematics for a different sample of
galaxies, using a single value of $\Upsilon_{\star}^{[3.6]}=0.5$ for
the disk component \citep{schombert14}.  They found an extremely small
vertical scatter in the BTFr of $\sigma=0.1$ dex.  This motivated us
to adopt a single mass--to--light ratio of
$\Upsilon_{\star}^{[3.6]}=0.5$ as one of the methods for estimating
the stellar mass of our sample galaxies.  We adopt an error on the
stellar mass-to-light ratio equal to $\epsilon_{\Upsilon_{\star}^{[3.6]}=0.5}=0.07$ 
dex as reported by \citet{schombert14}, and calculate the fractional error on the stellar
mass according to Eq.  \ref{stmasserreq}, with the magnitude error to 
be the mean error in the 3.6 $\mu$m apparent magnitudes, equal to $\epsilon_{m}=0.08$
mag.

\begin{table*}
\begin{tabular}{l r@{$\pm$}l r@{$\pm$}l r@{$\pm$}l r@{$\pm$}l r@{$\pm$}l r@{$\pm$}l}
\hline
Name & \multicolumn{2}{c}{$M_{\star,1}$}&\multicolumn{2}{c}{$M_{\star,2}$}&\multicolumn{2}{c}{$M_{\star,3}$}&\multicolumn{2}{c}{$M_{\star,4}$}&\multicolumn{2}{c}{$M_{atom}$}&\multicolumn{2}{c}{$M_{mol}$}  \\
&\multicolumn{2}{c}{$10^9 M_{\odot}$}&\multicolumn{2}{c}{$10^9 M_{\odot}$}&\multicolumn{2}{c}{$10^9 M_{\odot}$ }&\multicolumn{2}{c}{$10^9 M_{\odot}$}&\multicolumn{2}{c}{$10^9 M_{\odot}$ }&\multicolumn{2}{c}{$10^9 M_{\odot}$}\\
\hline                                                                                              
NGC 55 	    &1.6 &  0.9 &  1.3 & 0.8   & 2.2  &0.9  &   2.4    &1.0  &1.9  &0.01   &0.17   &0.05\\
NGC 224$^{a}$	&37.2 &  14  &  --  & --   & 83   &35   &  87      &35   &5.8  &0.68   &0.10   &0.03\\
NGC 247	    &1.1 &  0.2 &  1.5 & 0.9   & 3.4  &1.4  &   3.7    &1.5  &2.4  &0.17   &0.002  &0.0007\\
NGC 253	    &26.0 &  15  &  23  & 14    & 53   &22   &  57      &23   &2.9  &0.17   &2.56   &0.76\\
NGC 300	    &0.8 &  0.5 &  1.1 & 0.6   & 2.0  &0.8  &   2.1    &0.8  &2.2  &0.07   &0.05   &0.01\\
NGC 925	    &6.3 &  2.9 &  4.5 & 2.7   & 8.3  &3.5  &   9.0    &3.6  &7.2  &0.50   &0.28   &0.08\\
NGC 1365	&63.4 &  36.3&  49  & 29    & 86   &36   &  94      &38   &17   &0.57   &22.1   &6.64\\
NGC 2366	&0.01 &  0.01&  0.1 & 0.06  & 0.1  &0.06 &     0.1  &0.06 &1.1  &0.06   &0.01   &0.00\\
NGC 2403	&1.4 &  0.8 &  2.5 &  1.5  & 5.4  &2.2  &    6.0   &2.4  &3.6  &0.18   &0.63   &0.18\\
NGC 2541	&1.0 &  0.5 &  1.2 &  0.7  & 2.7  &1.1  &    2.9   &1.2  &6.3  &0.33   &0.03   &0.01\\
NGC 2841	&46.8 &  42.2&  50  &  30   & 101  &42   &   111    &44   &12   &1.02   &1.55   &0.46\\
NGC 2976	&0.7 &  0.4 &  0.8 &  0.5  & 1.7  &0.7  &    1.9   &0.8  &0.2  &0.009  &0.07   &0.02\\
NGC 3031	&38.5 &  24.8&  25  &  15   & 55   &23   &   57     &23   &3.9  &0.36   &0.45   &0.13\\
NGC 3109	&0.04 &  0.01&  0.07&  0.04 & 0.1  &0.05 &     0.1  &0.06 &0.7  &0.05   &0.01   &0.007\\
NGC 3198	&7.7 &  4.6 &  8.3 &  4.9  & 16   &6.8  &    18    &7.3  &12.  &0.74   &0.50   &0.15\\
IC 2574 	&0.05 &  0.02&  0.2 &  0.1  & 0.7  &0.3  &    0.8   &0.3  &1.9  &0.10   &0.05   &0.001\\
NGC 3319	&2.2 &  1.2 &  2.4 &  1.4  & 4.2  &1.7  &    4.6   &1.8  &5.1  &0.28   &0   &0\\
NGC 3351	&22.4 &  12.7&  16  &  9.7  & 29   &12   &   32     &13.  &2.1  &0.11   &1.40   &0.42\\
NGC 3370	&3.6 &  2.1 &  7.6 &  4.5  & 17   &7.2  &    18    &7.6  &3.9  &0.27   &0.57   &0.17\\
NGC 3621	&8.4 &  6.1 &  6.1 &  3.6  & 12   &5.3  &    14    &5.7  &13   &0.74   &1.78   &0.53\\
NGC 3627	&16.1 &  9.2 &  24  &  14   & 45   &19   &   50     &20   &1.4  &0.09   &5.76   &1.72\\
NGC 4244	&0.7 &  0.3 &  1.1 &   0.6 & 2.6  &1.1 &    2.8 &  1.1   &2.9  &0.16   &0      & 0\\
NGC 4258	&23.2 &  20.8&  25  &  15   & 52   &22   &   54     &22   &7.7  &0.58   &1.82   &0.54\\
NGC 4414	&34.4 &  19.7&  32  &  19   & 59   &25   &   65     &26   &7.2  &0.54   &4.50   &1.35\\
NGC 4535	&23.3 &  17.1&  23  &  13   & 42   &18   &   47     &19   &6.5  &0.25   &5.47   &1.64\\
NGC 4536	&15.6 &  11.9&  11  &  6.7  & 24   &10   &   27     &11   &5.8  &0.24   &2.41   &0.72\\
NGC 4605	&1.8 &  1   &  1.4 &  0.8  & 3.0  &1.2  &    3.4   &1.3  &0.5  &0.03   &0.11   &0.03\\
NGC 4639	&14.1 &  4.7 &  10  &  6.1  & 19   &8.2  &    20    &8.4  &2.2  &0.15   &0.24   &0.07\\
NGC 4725	&46.2 &  26.4&  33  &  19   & 58   &24   &   60     &24   &5.3  &0.19   &0.01   &0.00\\
NGC 5584	&2.9 &  4.2 &  7.2 &  4.2  & 11   &4.7  &    12    &5.1  &2.6  &0.15   &0.91   &0.27\\
NGC 7331	&123.6 &  70.4&  53  &  31   & 107  &45   &   118    &47   &12   &0.94   &11     &3.31\\
NGC 7793	&2.4 &  1   &  1.5 &  0.9  & 2.9  &1.2  &    3.2   &1.3  &1.4  &0.09   &0.45   &0.13\\
\hline
\end{tabular}
\caption{Stellar and gas masses of the sample galaxies.
Column (1): galaxy name;
Column (2-5): stellar mass, estimated with different methods: 1-- using SED--fitting,
 2-- using dynamical $\Upsilon_{\star}^{Dyn, K}=0.29$, 3-- using $\Upsilon_{\star}$ as a function of [3.6]-[4.5]
 colour, 4-- using constant $\Upsilon_{\star}=0.5$;
Column (6): total mass of the atomic gas, including contribution of helium and heavier elements;
Column (7): total mass of the molecular gas, including contribution of helium and heavier elements;
 a) we remind that there is no data available for NGC 224 in the K--band. Therefore it lacks a 
stellar mass estimate based on the second method.
}
\label{tbl_masses}
\end{table*}

\subsection{A comparison between stellar mass--to--light ratios} \label{stmasscomp}


\begin{figure}
\begin{center}
\includegraphics[scale=0.60]{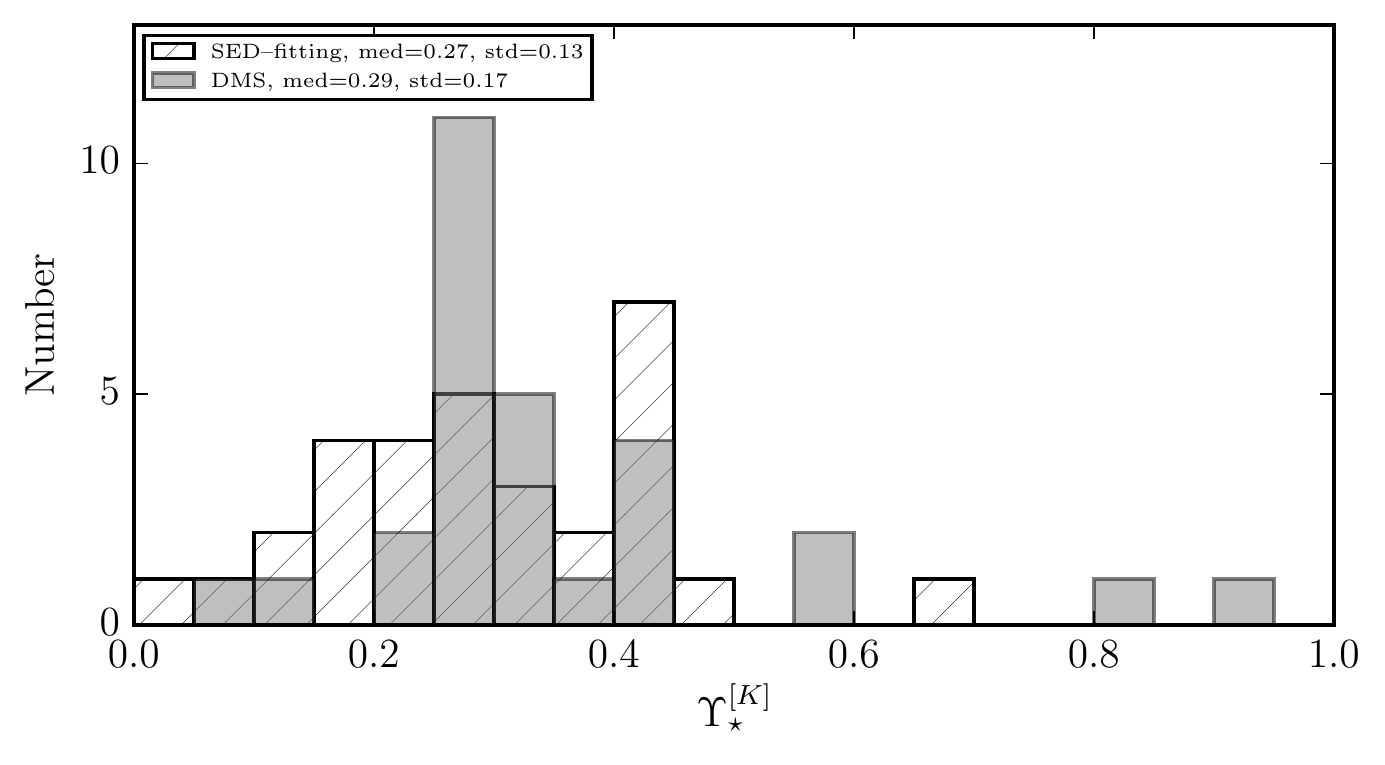}
\caption{Comparison of the distribution of stellar mass-to-light
  ratios for the $K$--band from the DiskMass Survey for a sample of 30
  face-on galaxies (dark shade) and from the SED-fitting for our
  sample (hatched). Distributions have almost the same median with
  a difference of only 0.02.}
\label{fig_seddm}
\end{center}
\end{figure}

\begin{figure}
\begin{center}
\includegraphics[scale=0.60]{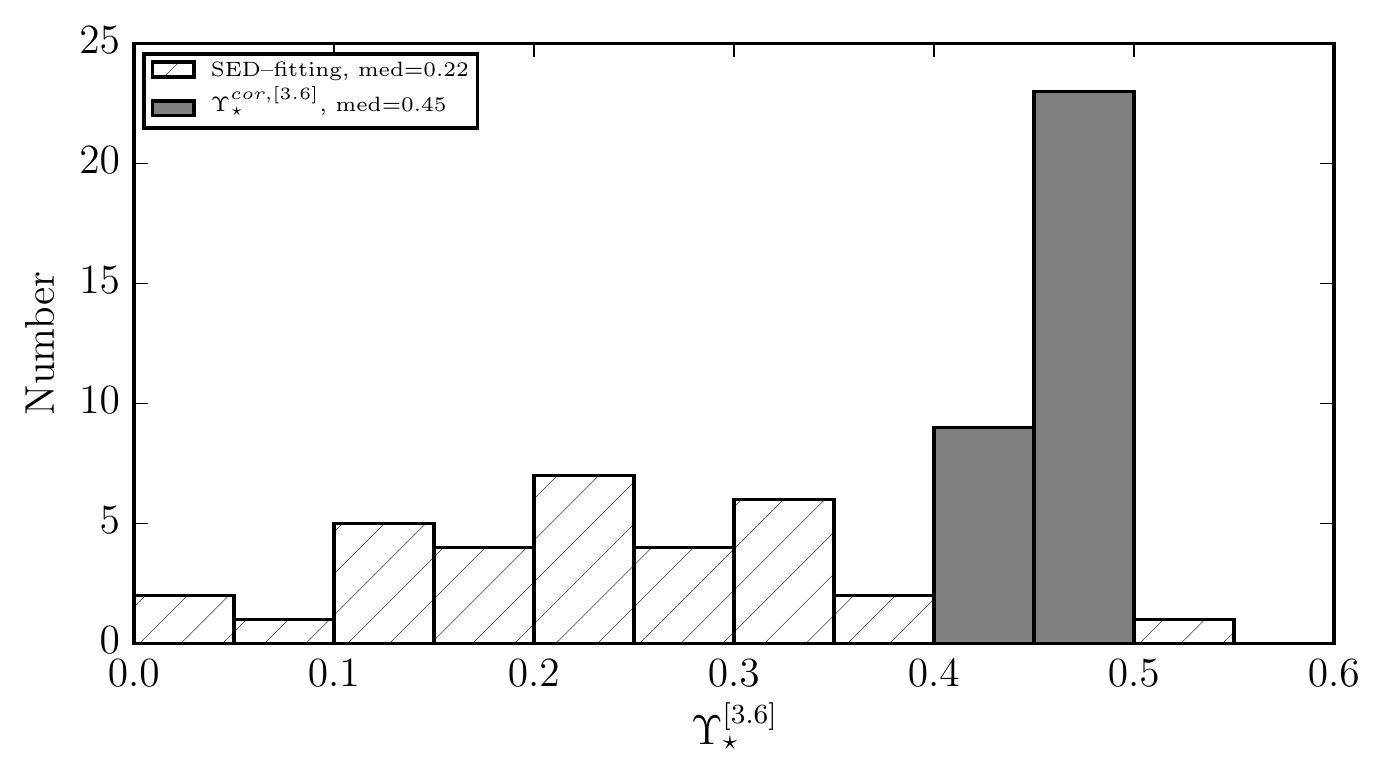}
\caption{ Comparison of the distribution of stellar mass-to-light
  ratios at 3.6 $\mu$m as a function of colour (Method 3) in dark shade and from the
  SED-fitting (Method 1) in hatched area. The distribution of
  $\Upsilon_{\star}^{SED,[3.6]}$ is much broader than the distribution
  of $\Upsilon_{\star}^{[3.6], cor}$.}
\label{fig_sedcor}
\end{center}
\end{figure}

\begin{figure*}
\begin{center}
\includegraphics[scale=0.70]{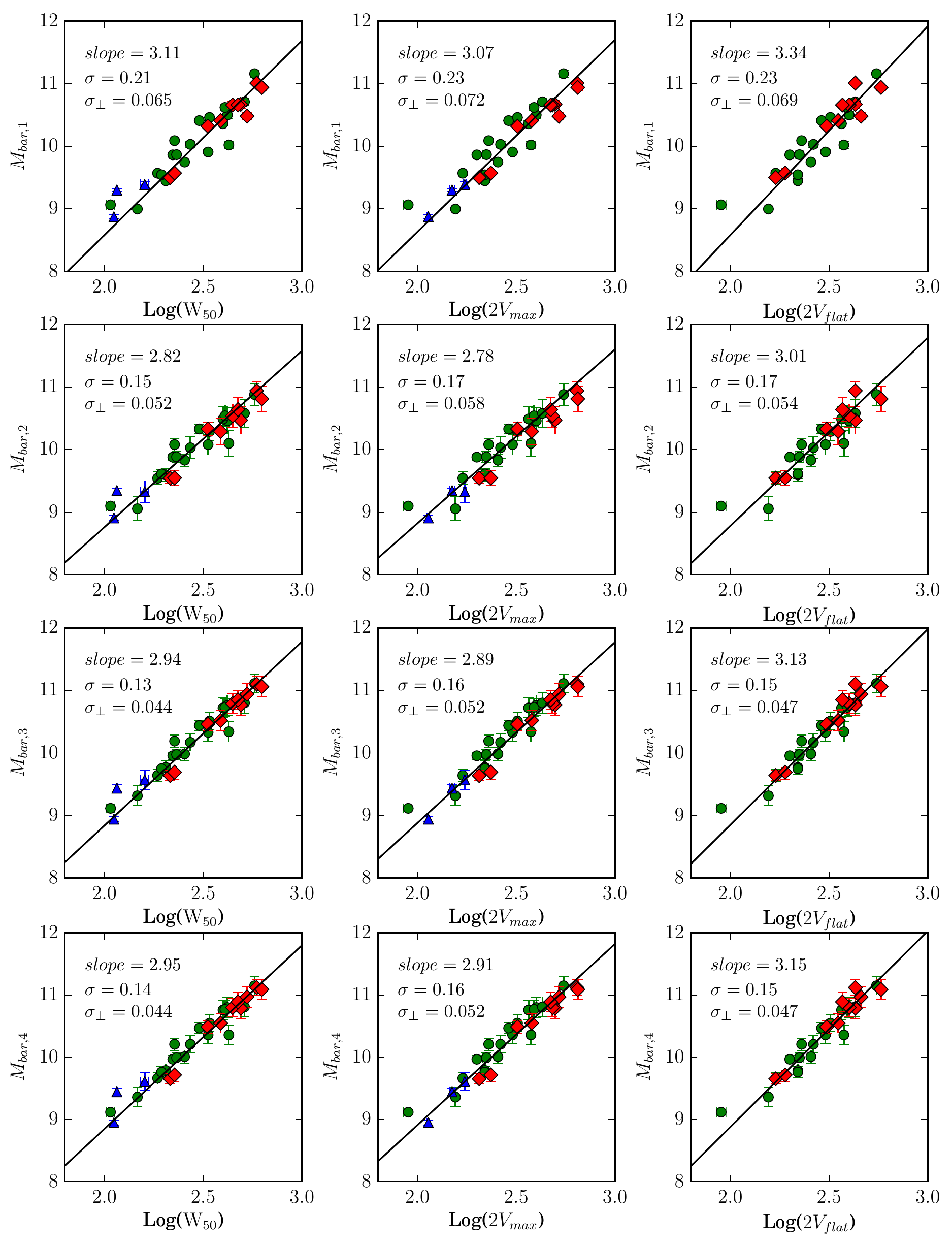}
\caption{ The BTFrs based on the different rotational velocity
  measures and using different stellar mass estimates.  From top to
  bottom 1. using SED--fitting; 2. using dynamical mass--to--light ratio
  calibration $<\Upsilon_{\star}^{Dyn, K}>=0.29$; 3. using
  $\Upsilon_{\star}^{[3.6], cor}$ as a function of $[3.6]-[4.5]$
  colour; 4. using constant $\Upsilon_{\star}^{[3.6]}=0.5$. The
  best-fit models are shown with solid lines.  Green symbols show flat
  rotation curves ($V_{max} = V_{flat}$), and red symbols indicate
  galaxies with declining rotation curves ($V_{max} >V_{flat}$). Blue
  symbols indicate galaxies with rising rotation curves
  ($V_{max} < V_{flat}$). These galaxies were not included when
  fitting the model.}
\label{fig_allbtf}
\end{center}
\end{figure*}

The four different methods from the previous subsections have
demonstrated that stellar masses of spiral galaxies can not be
estimated straightforwardly with a single prescription.  The resulting
stellar masses derived with these different methods are summarised in
Table \ref{tbl_masses}.  Here we conclude with comparisons between the
derived stellar mass--to--light ratios.







\begin{figure}
\begin{center}
\includegraphics[scale=0.7]{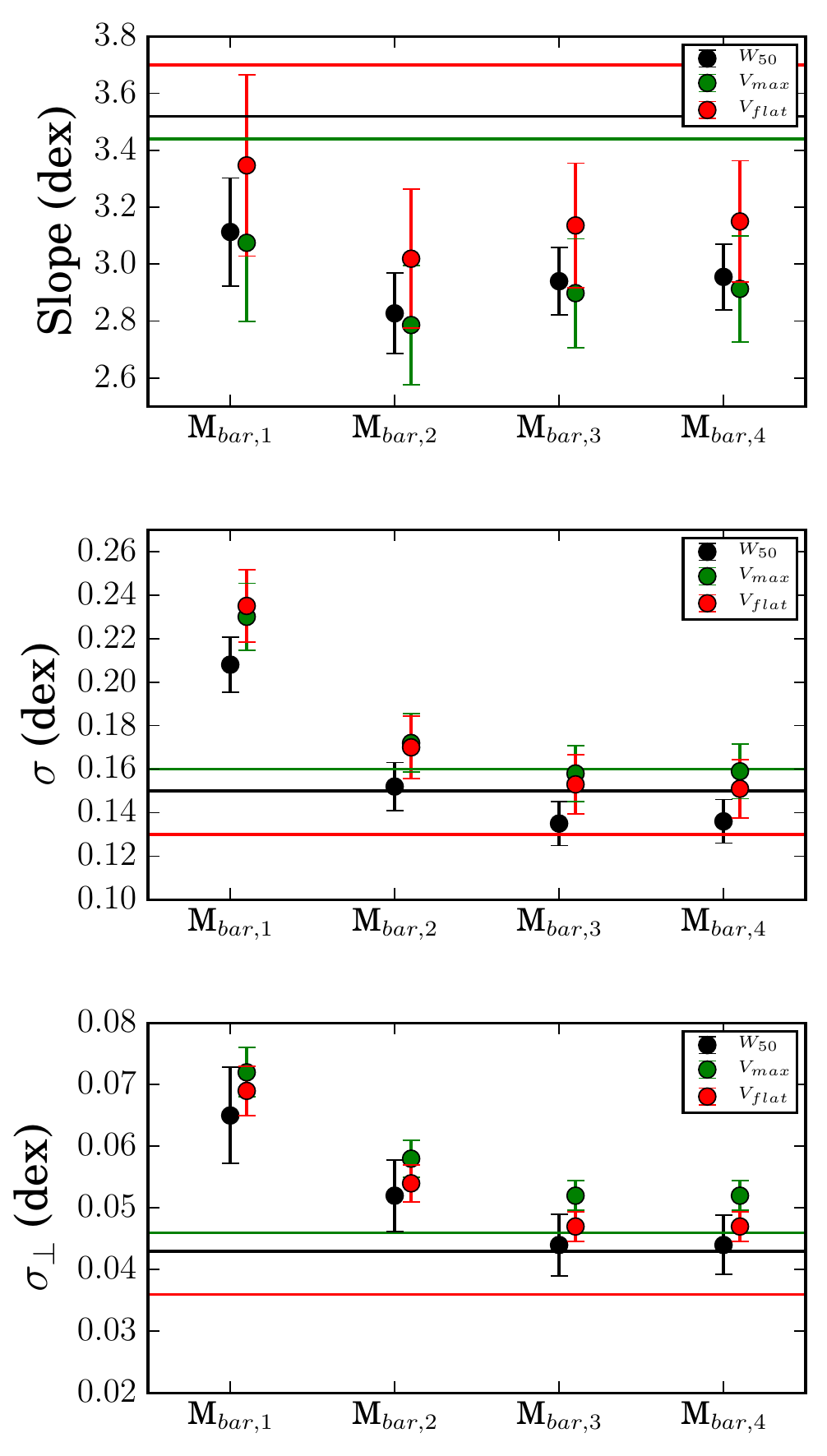}
\caption{ The slope, vertical scatter and tightness of the
  BTFrs. Black symbols indicate the values for the relation based on
  $W_{50}$ as a rotational velocity measure, green on $V_{max}$ and
  red on $V_{flat}$.  The values are presented for the BTFrs, using
  different stellar mass estimates: 1. SED--fitting; 2. dynamical
  $<\Upsilon_{\star}^{Dyn, K}>=0.29$; 3.
  $\Upsilon_{\star}^{[3.6], cor}$ as a function of $[3.6]-[4.5]$
  colour; 4. constant $\Upsilon_{\star}^{[3.6]}=0.5$. The solid lines
  show the slope, scatter and tightness of the 3.6 $\mu$m
  luminosity--based TFr based on the different rotational velocity
  measures.}
\label{fig_mbartfr}
\end{center}
\end{figure}

\begin{figure}
\begin{center}
\includegraphics[scale=0.7]{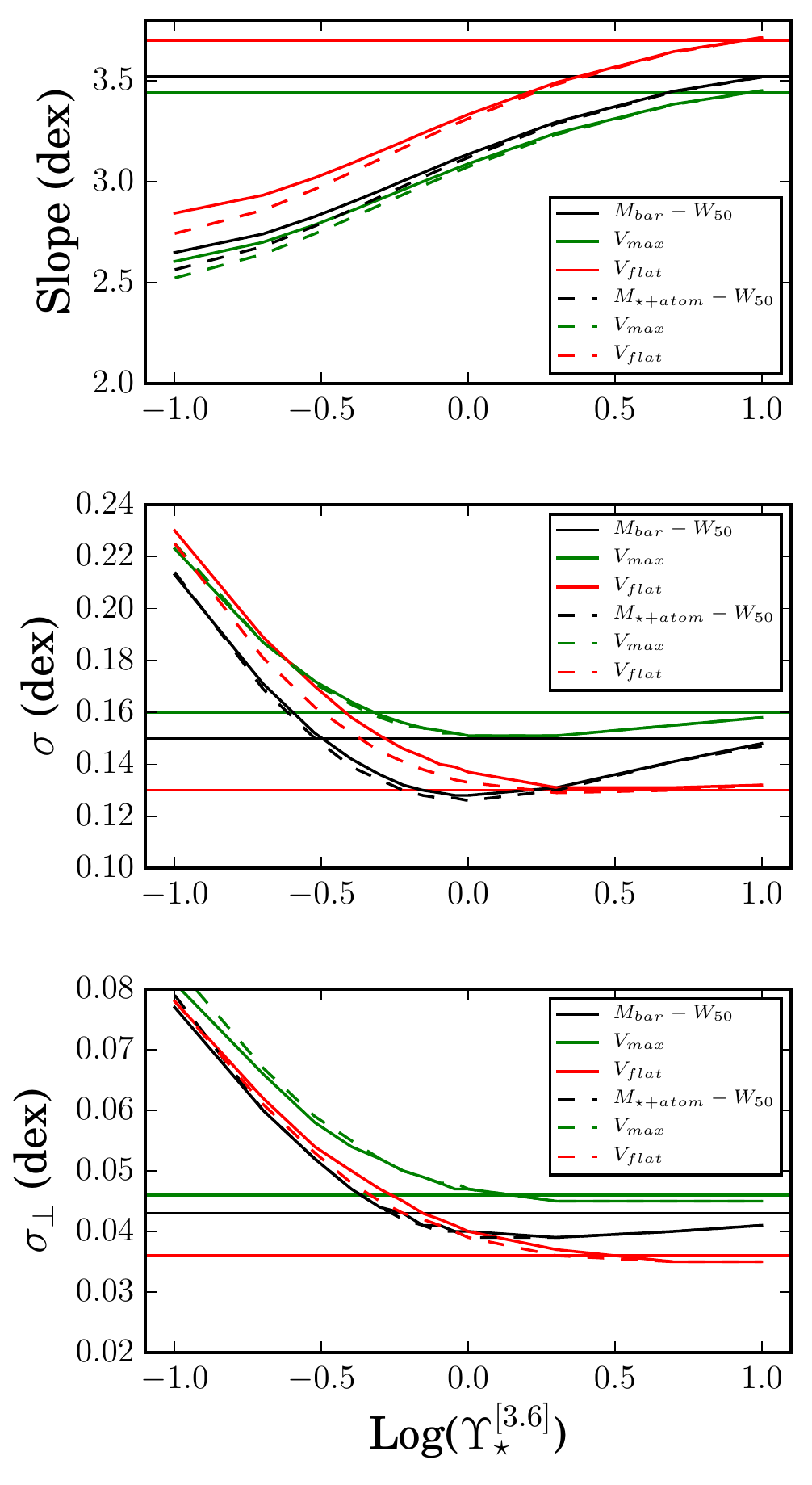}
\caption{The slope, vertical scatter and tightness of the BTFrs as a
  function of the mass--to-light ratio in the 3.6 $\mu$m band. Black lines
  indicate the values for the relation based on $W_{50}$ as a
  rotational velocity measure, green on $V_{max}$ and red on
  $V_{flat}$.  The solid lines show the trends for the BTFr
  ($M_{\star}+M_{atom}+M_{mol}$).  The dashed lines show the trends
  for $M_{\star}+M_{atom}$.  The solid horizontal lines show the
  slope, scatter and tightness of the 3.6 $\mu$m luminosity--based TFr
  based on the different rotational velocity measures. }
\label{fig_mls}
\end{center}
\end{figure}

We find that the stellar mass--to--light ratios obtained with the
SED--fitting cover a wide range of values between 0.04 and 0.67 for
the $K$--band and from 0.03 to 0.52 in the 3.6 $\mu$m band.  Such a
large scatter in the bands which are considered to have more or less
the same mass--to--light ratio for all galaxies, can be driven by the
measurement errors and model uncertainties. Indeed, it is very
complicated to assign a single mass--to--light ratio even within a
galaxy, as spirals tend to have various components, such as a bulge,
disk and spiral arms. Therefore, gradients in the mass--to--light
ratio are likely to be present within a galaxy, indicating the
differences in IMF and in star formation histories.  However, in our
analysis we do not consider radial trends in mass--to--light ratios,
which may also be a reason for the large scatter and uncertainties in
$\Upsilon_{\star}^{SED, K/[3.6]}$. Interestingly, the values of the
dynamical mass--to--light ratios in the $K$--band from the DMS are
also spread over a wide range between 0.06 and 0.94.


Figure \ref{fig_seddm} presents the comparison between the
distribution of $\Upsilon_{\star}^{K}$ from the DMS and from the
SED--fitting for different but representative samples of spiral
galaxies.  Remarkably, these distributions are very similar with a
difference in the median of only 0.01, even though the values are
measured using different methods for different samples.  Furthermore,
a comparison between $\Upsilon_{\star}^{SED,[3.6]}$ from the
SED--fitting and $\Upsilon_{\star}^{[3.6], cor}$ as a function of the
[3.6]--[4.5] colour is shown in Figure \ref{fig_sedcor}. While the
$\Upsilon_{\star}^{SED,[3.6]}$ is ranging from 0.03 to 0.52, the
$\Upsilon_{\star}^{[3.6], cor}$ is spread over a much narrower range
from 0.44 to 0.49. The range of $\Upsilon_{\star}^{[3.6], cor}$ is
driven by the difference between the uncorrected 3.6 $\mu$m flux and
the flux corrected for non--stellar contamination, which can be
significant in spiral galaxies.


\section{A comparison of Baryonic Tully--Fisher relations} \label{btfrscomp}

In this section we present the BTFrs based on different rotational
velocity measures ($W_{50}$, $V_{max}$ and $V_{flat}$) and using
different stellar mass estimates (see Section \ref{stmass}) in order
to study how the slope, scatter and tightness of the BTFr depend on
these parameters.

We calculate the baryonic mass of a galaxy as the sum of the
individual baryonic components: stellar mass, atomic gas mass and
molecular gas mass, as listed in Table \ref{tbl_masses}:

\begin{equation} \label{mbar}
M_{bar,m} =M_{\star,m}+M_{atom}+M_{mol},
\end{equation}

\noindent
where $M_{\star,m}$ is one of four stellar masses, estimated with the
four different methods ($m=1,2,3,4$).  We further calculate the error
on the baryonic mass by applying a full error propagation calculation:

\begin{equation}
\Delta M_{bar,m}=\sqrt{\Delta M_{\star,m}^2+\Delta M_{atom}^2+\Delta M_{mol}^2},
\end{equation}

\noindent
The derivation of $\Delta M_{\star,m}$, $\Delta M_{atom}$ and
$\Delta M_{mol}$ is described in Sections \ref{gassection} \&
\ref{stmass}.

\begin{table*}
\begin{tabular}{lcccccc}
\hline
 $M_{bar}$&\multicolumn{3}{c}{Slope}&\multicolumn{3}{c}{Zero point}\\
\hline
  &$W_{50}$& $V_{max}$ &$V_{flat}$ &$W_{50}$& $V_{max}$ &$V_{flat}$\\
\hline
$M_{bar,1}$ & 3.11$\pm$0.19 & 3.07$\pm$0.27 &  3.34$\pm$0.31 & 2.36$\pm$0.48  &2.49$\pm$0.70   &1.90$\pm$0.80 \\    
$M_{bar,2}$ & 2.82$\pm$0.14 & 2.78$\pm$0.20 &  3.01$\pm$0.24 & 3.12$\pm$0.36  &3.26$\pm$0.53   &2.76$\pm$0.61 \\    
$M_{bar,3}$&  2.94$\pm$0.11 & 2.89$\pm$0.19 &  3.13$\pm$0.21 & 2.96$\pm$0.30  &3.10$\pm$0.48   &2.59$\pm$0.54 \\   
$M_{bar,4}$&  2.95$\pm$0.11 & 2.91$\pm$0.18 &  3.15$\pm$0.21 & 2.95$\pm$0.29  &3.09$\pm$0.47   &2.58$\pm$0.53 \\   
\hline
\end{tabular}
\caption{The statistical properties of the BTFrs. 
  Column (1): Baryonic mass of a galaxy with different stellar mass estimations;
  Column (2)-Column(4): slopes of the BTFrs based on $W_{50}$, $V_{max}$ and $V_{flat}$;
  Column (5)-Column(7): zero points of the TFrs based on $W_{50}$, $V_{max}$ and $V_{flat}$}
\label{tbl_slope}
\end{table*} 

\begin{table*}
\begin{tabular}{lcccccc}   
\hline
&\multicolumn{3}{c}{vertical scatter ($\sigma$)}&\multicolumn{3}{c}{tightness ($\sigma_{\perp}$)}\\
\hline
  &$W_{50}$& $V_{max}$ &$V_{flat}$ &$W_{50}$& $V_{max}$ &$V_{flat}$\\
  \hline
$M_{bar,1}$& 0.21$\pm$0.01 & 0.23$\pm$0.02 & 0.23$\pm$0.02 & 0.065$\pm$0.008 & 0.072$\pm$0.004 & 0.069$\pm$0.004   \\    
$M_{bar,2}$& 0.15$\pm$0.01 & 0.17$\pm$0.01 & 0.17$\pm$0.01 & 0.052$\pm$0.006 & 0.058$\pm$0.003 & 0.054$\pm$0.003   \\    
$M_{bar,3}$& 0.13$\pm$0.01 & 0.16$\pm$0.01 & 0.15$\pm$0.01 & 0.044$\pm$0.005 & 0.052$\pm$0.002 & 0.047$\pm$0.002   \\   
$M_{bar,4}$& 0.14$\pm$0.01 & 0.16$\pm$0.01 & 0.15$\pm$0.01 & 0.044$\pm$0.006 & 0.052$\pm$0.002 & 0.047$\pm$0.002   \\   
\hline
\hline
\end{tabular}
\caption{The statistical properties of the BTFrs (continued).
  Column (1): Baryonic mass of a galaxy with different stellar mass estimations;
  Column (2)-Column(4): scatters of the BTFrs based on $W_{50}$, $V_{max}$ and $V_{flat}$;
  Column (5)-Column(7): tightnesses of the BTFrs based on $W_{50}$, $V_{max}$ and $V_{flat}$.}
\label{tbl_scatter}
\end{table*}  

\begin{figure*}
\includegraphics[scale=0.70]{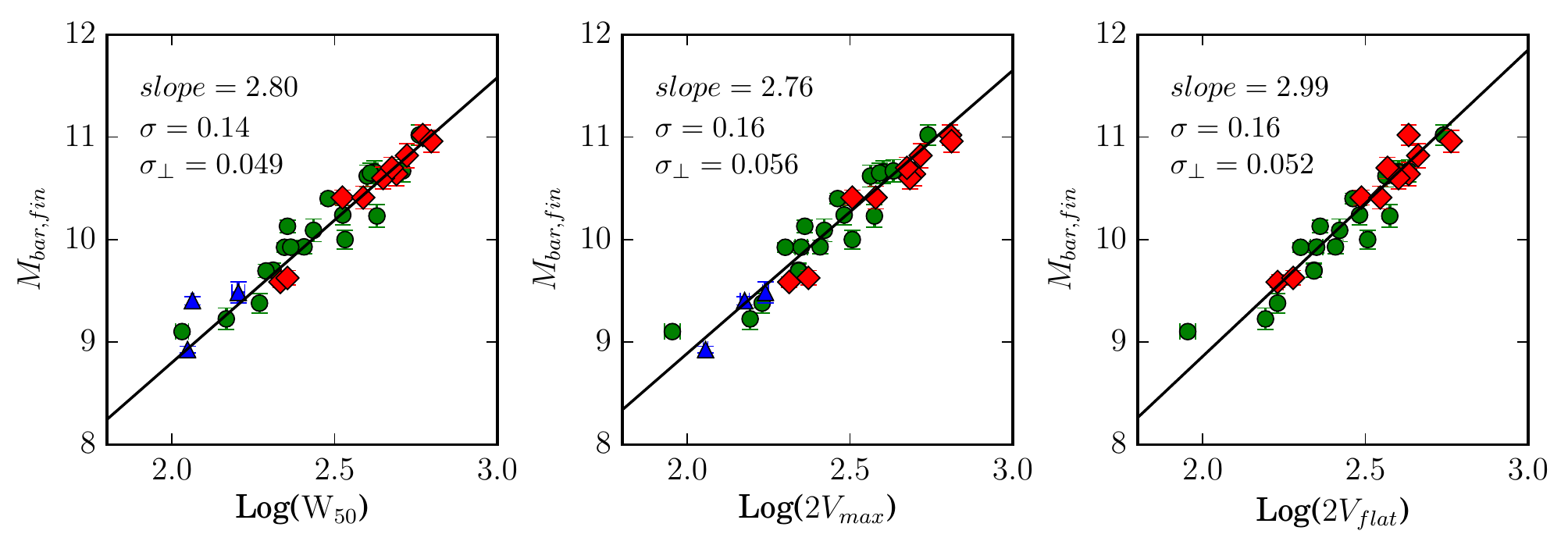}
\caption{The final BTFr based on the three velocity measures ($W_{50}$,
  $V_{max}$ and $V_{flat}$) with the baryonic mass calculated as
  $M_{bar,fin}=M_{\star,fin}+M_{atom}+M_{mol}$. The best-fit models
  are shown with solid lines.  Green symbols show galaxies with flat rotation curves
  ($V_{max} = V_{flat}$), and red symbols indicate galaxies with
  declining rotation curves ($V_{max} >V_{flat}$). Blue symbols
  indicate galaxies with rising rotation curves
  ($V_{max} < V_{flat}$). These galaxies were not included when
  fitting the model. }
\label{fig_fintfr}
\end{figure*}

Consequently, we obtain 12 BTFrs for which we measure slope, scatter
and tightness.  To be able to perform a fair comparison with the
statistical properties of the luminosity--based TFr, we calculate the
above mentioned values of scatter and tightness in the BTFrs in the
same manner as described in \citetalias{pon17}.  All 12 relations are
shown in Figure \ref{fig_allbtf} together with the best fit models of
the form $log M_{bar}=a\times log V_{rot}+b$.

First, we perform an orthogonal fit to the data points, where the
best--fit model minimises the orthogonal distances from the data
points to the model. We use the python implementation of the $BCES$
fitting method \citep{akritas96,nemmen12}, which allows to take
correlated errors in both directions into account.  Moreover, with
this method we assign less weight to outliers and to data points with
large error bars. Subsequently, we calculate the vertical scatter
$\sigma$ and the perpendicular tightness $\sigma_{\perp}$ of each
relation as described in \citetalias{pon17}.

\begin{figure}
\begin{center}
\includegraphics[scale=0.85]{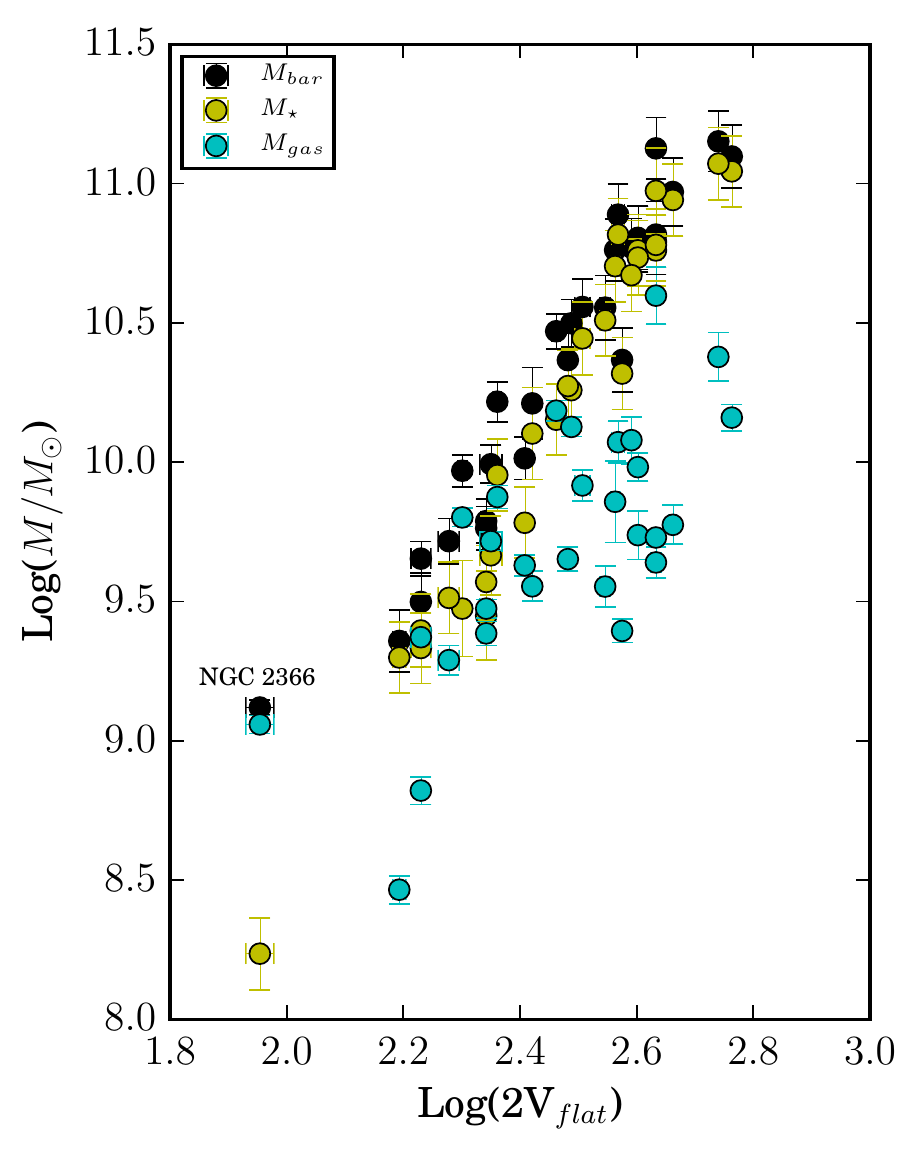}
\caption{The final choice BTFr based on $V_{flat}$ is shown with the
  black symbols. The stellar component is shown with the yellow symbols and
  the gaseous component with the cyan symbols. }
\label{fig_finbtfr}
\end{center}
\end{figure}

Figure \ref{fig_mbartfr} shows the slope, scatter and tightness of the
BTFrs for different rotational velocity measures and using different
stellar mass estimates. We find that the BTFr with the stellar mass
estimated from the SED--fitting ($M_{bar,1}$) shows the largest
observed vertical and perpendicular scatter. Next, the BTFr with the
stellar mass based on the DMS--motivated dynamical mass--to--light
ratio estimate of $<\Upsilon_{\star}^{Dyn,K}>=0.29$ ($M_{bar,2}$),
demonstrates somewhat less scatter and appears to be tighter. However,
the tightest BTFrs with the smallest vertical scatter are those BTFrs
with higher stellar mass--to--light ratios ($M_{bar,3}$ and
$M_{bar,4}$). Moreover, all the BTFrs demonstrate a shallower slope
and a larger scatter, and are less tight compared to the 3.6 $\mu$m
luminosity--based TFr \citepalias{pon17}. This result is contrary to
previous studies \citep{mcgaugh00,mcgaugh05}, since it suggests that
inclusion of the gas mass does not help to tighten the TFr.  Instead,
it introduces additional scatter, especially for the lower stellar
mass-to-light ratios.


We performed a test by assigning different mass--to--light ratios in
the 3.6 $\mu$m band for our sample. We vary mass--to--light ratios
from 0.1 to 10, but assign the same value to all galaxies. From Figure
\ref{fig_mls} it is also clear, that increasing the mass--to--light
ratio helps to reduce the vertical scatter and improve the tightness
of the BTFr, suggesting that the scatter in the BTFr is introduced by
the gaseous component.  From Figure \ref{fig_mls} it is also clear
that the contribution of the molecular gas component does not
significantly affect the statistical properties of the BTFr.


The other important result from our study is that, independent of the
stellar mass estimate, each BTFr shows a smaller scatter and improved
tightness when based on $W_{50}$ as a rotational velocity
measure. This result is also in contradiction with theoretical
hypotheses concerning the origin of the TFr, being a relation between
the baryonic mass of a galaxy and that of its host dark matter
halo. Only $V_{flat}$ can properly trace the gravitational potential
of a dark matter halo, because it is measured in the outskirts of the
extended H{\sc i} disk where the potential is dominated by the dark
matter halo.  However, it is also important to note that the scatter
and tightness of the BTFr based on $W_{50}$ and on $V_{flat}$ are
consistent within their error.  Tables \ref{tbl_slope} \&
\ref{tbl_scatter} summarise the statistical properties of the BTFrs.

\section{Our adopted Baryonic Tully--Fisher relation}\label{adoptedbtfr}

\begin{figure}
\begin{center}
\includegraphics[scale=0.9]{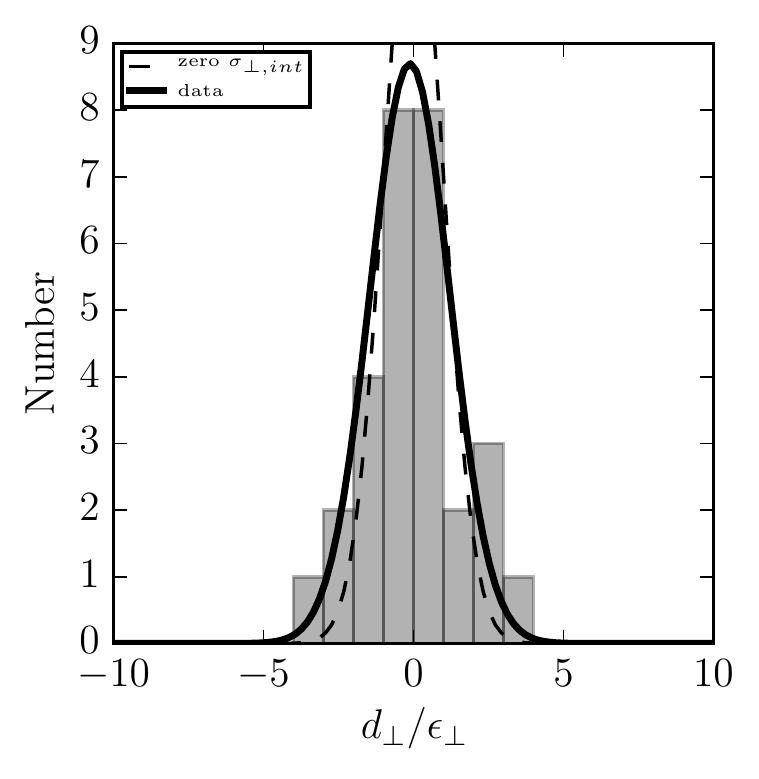}
\caption{Histogram of the perpendicular distances from the data points
  to the line ($d_{\perp,}$) in $M_{bar,fin}$--$V_{flat}$ relation,
  normalised by the perpendicular errors.  The standard normal
  distribution that would be expected for a zero intrinsic tightness
  is shown with dashed line. The best--fit to the data, weighted by the
  Poisson errors, is shown with the solid line with a standard
  deviation of 1.33 $\pm$ 0.2. }
\label{fig_finhist}
\end{center}
\end{figure}

\begin{figure*}
\begin{center}
\includegraphics[scale=0.6]{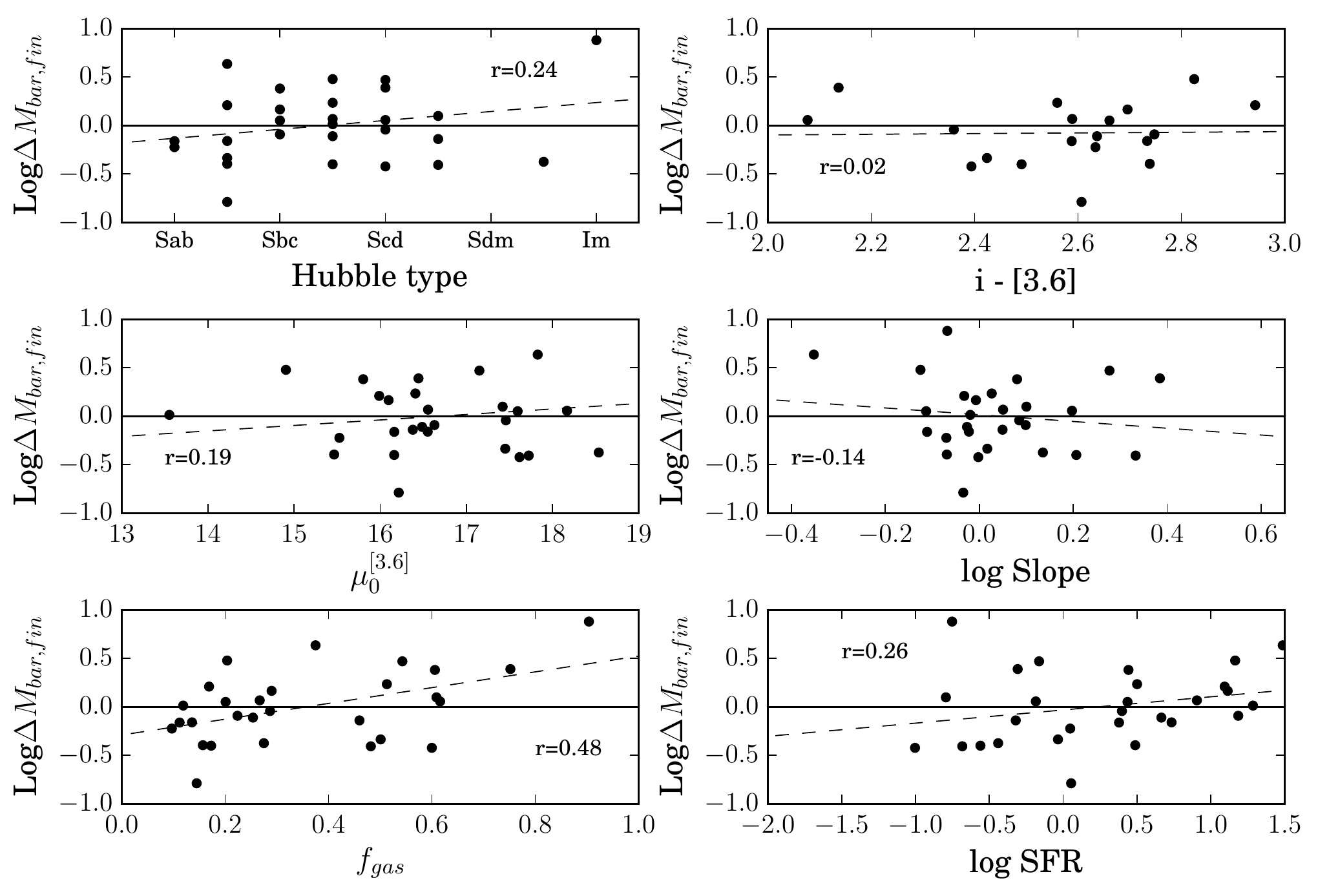}
\caption{Residuals of the $M_{bar,fin}$--$2V_{flat}$ relation as a
  function of global galactic properties.  $r$ is Pearson's
  correlation coefficient. }
\label{fig_resbarfin}
\end{center}
\end{figure*}

As was described in the previous sections, the choice of the stellar
mass--to--light ratio is not straightforward and requires an
evaluation of estimates based on various methods, e.g. from stellar
population modelling, or from dynamical modelling.  Interestingly,
from our SED--fitting and from the dynamical estimate of
$\Upsilon_{\star}^{K}$ from the DMS, we obtained the same median of
$<\Upsilon_{\star}^{K}> \sim 0.3$ (Figure \ref{fig_seddm}) for
galaxies that cover similar morphological types.  However, the method
of estimating the mass--to--light ratio as a function of the
[3.6]--[4.5] colour gives a somewhat larger mass--to--light ratios
with a much smaller scatter; $\Upsilon_{\star}^{[3.6], cor}$ lies in
the range between 0.44 and 0.49.

From Section \ref{stmasscomp} we conclude that the individual
mass--to--light ratios from the SED--fitting are not applicable to
our galaxies, as their values show a large scatter, including
unrealistically low mass--to--light ratios for several galaxies.  From
our study of the statistical properties of the individual BTFrs we
also can not draw certain conclusions regarding which stellar
mass--to--light ratio to adopt, as the scatter in the BTFr for each
case is driven by the gas component (see Section
\ref{btfrscomp}). Therefore, for a more detailed study of the BTFr we
adopt an intermediate value of the mass--to--light ratio in 3.6 $\mu$m
band between the high values coming from methods 3 \& 4 and the low
values coming from methods 1 \& 2. Hence, we adopt the final
mass--to--light ratio of $\Upsilon_{\star}^{fin,[3.6]}=0.35$, which we
assign to all galaxies in our sample.


We calculate the baryonic mass of our sample galaxies $M_{bar,fin}$,
according to Eq. \ref{mbar} with the stellar mass measured as
$M_{\star,fin}=\Upsilon_{\star}^{fin,[3.6]} \cdot
L_{[3.6]}(L_{\odot})$.
Figure \ref{fig_fintfr} shows our final BTF relations based on the
three velocity measures $W_{50}$, $V_{max}$ and $V_{flat}$.  The
$M_{bar,fin}$--$V_{flat}$, BTFr according to our fit, can be described
as

\begin{equation}\label{mbarfin}
M_{bar,fin}=(2.99\pm0.22)  \cdot log(2V_{flat}) + 2.88\pm 0.56.
\end{equation}
Eq. \ref{mbarfin} describes the relation with an observed vertical
scatter of $\sigma=0.16\pm 0.1$ dex and a tightness of
$\sigma_{\perp, obs}=0.052 \pm 0.013$ dex. These results are
consistent with recent studies of the vertical \citep{lelli16} and
perpendicular \citep{manolistfr} scatters of the BTFr, but are
somewhat larger compared to the 3.6 $\mu$m luminosity--based TFr \citepalias{pon17}.  
Importantly, our slope and scatter are consistent with the mean value 
for these parameters as a function of BTFr sample size \citep{sorceguo16}.
The contributions from the stellar and gaseous components to the BTFr
separately are shown in Figure \ref{fig_finbtfr}

Furthermore, we investigate the intrinsic tightness of the BTFr. We
are focusing on the tightness and not on the vertical scatter of the
relation, because the tightness is a slope independent measure and
should be used as a possible constraint on theories of galaxy
formation and evolution.  We compare the perpendicular distances
$d_{\perp,i}$ of the data points to the line, with the projected
measurement errors $\epsilon_{i}$ based on the error on the baryonic
mass ($\Delta M_{bar,i}$) and the error on the rotational velocity
($\epsilon_{V_{flat},i}$) (see \citetalias{pon17} for more details).
Figure \ref{fig_finhist} shows the histogram of
$d_{\perp,i}/\epsilon_{i}$. In the case of zero intrinsic scatter
$\sigma_{\perp}$, this histogram would follow the standard normal
distribution, shown with the dashed Gaussian curve in Figure
\ref{fig_finhist}.  However, it is clear that the distribution of
$d_{\perp,i}/\epsilon_{i}$, shown with the solid Gaussian curve, is
broader with a standard deviation of 1.33$\pm$0.2.  Consequently,
we can estimate the value of the intrinsic $\sigma_{\perp, int}$ as
follows:

\begin{equation}
\sigma_{\perp, int}=\sqrt{\sigma_{\perp, obs}^{2}-\sigma_{\perp,err}^{2}},
\end{equation}

\noindent
where $\sigma_{\perp,err}=0.045$ dex is the perpendicular scatter due
to the measurement uncertainties only.  Hence, we estimate the
$\sigma_{\perp, int} \sim 0.026 \pm 0.013$ dex, which is identical to the
$\sigma_{\perp, int}$ of the 3.6 $\mu$m luminosity--based TFr
\citepalias{pon17}.  Therefore, we can conclude that even if the BTFr
has a larger observed perpendicular scatter compared to the 3.6 $\mu$m
luminosity--based TFr, it is only due to the measurement uncertainties
because both relations have the same intrinsic perpendicular scatter
$\sigma_{\perp, int} \sim 0.026$ dex.

\subsection{Search for a 2$^{nd}$ parameter}

As was suggested by various authors \citep{aar84, rubin85}, the
vertical scatter in the luminosity--based TFr at optical wavelengths
can be decreased by invoking a second parameter. However, we
demonstrated in \citetalias{pon17} that the residuals of the 3.6
$\mu$m TFr do not correlate significantly with any of the galactic
properties and, therefore, we could not identify any second paramete
that could further reduce the scatter.  \citet{v01} reached
the same conclusion for the K-band $M_{K}$--$V_{flat}$ TFr constructed
with Ursa Major galaxies.  In this section we repeat this exercise for
the $M_{bar,fin}$--$V_{flat}$ relation and examine the nature of the
residuals along the fitted model line described by Eq. \ref{mbarfin}.
Figure \ref{fig_resbarfin} presents the residuals of the BTFr
(log$\Delta M_{bar}$) as a function of global galactic properties,
such as star formation rate, outer slope of the rotation curve,
central surface brightness, $i-[3.6]$ colour, and gas fraction. We
calculate $\Delta M_{bar}$ as $\Delta M_{bar}=M_{bar}/ M_{bar,model}$,
where $M_{bar,model}$ is described by Eq. \ref{mbarfin}.

To quantitatively describe the strength of the correlations we
calculate Pearson's correlation coefficients $r$ for each of the
relations. We find the largest $r=0.48$ for the correlation between
$\Delta M_{bar}$ and the total gas fraction
($f_{gas}=(M_{atom}+M_{mol})/M_{bar}$) and the smallest $r=0.02$ for
the correlation between $\Delta M_{bar}$ and $i-[3.6]$
colour. Eventhough the strength of neither correlations is sufficient
to identify a significant second parameter, a possible correlation
between $\Delta M_{bar}$ and $f_{gas}$ could explain why studies based
on gas-rich galaxies \citep{mcgaugh12,manolistfr} find generally
steeper slopes.

%


\section{Comparison with previous observational studies and theoretical results}\label{comp}


\subsection{Previous studies}

\begin{figure*}
\begin{center}
\includegraphics[scale=0.5]{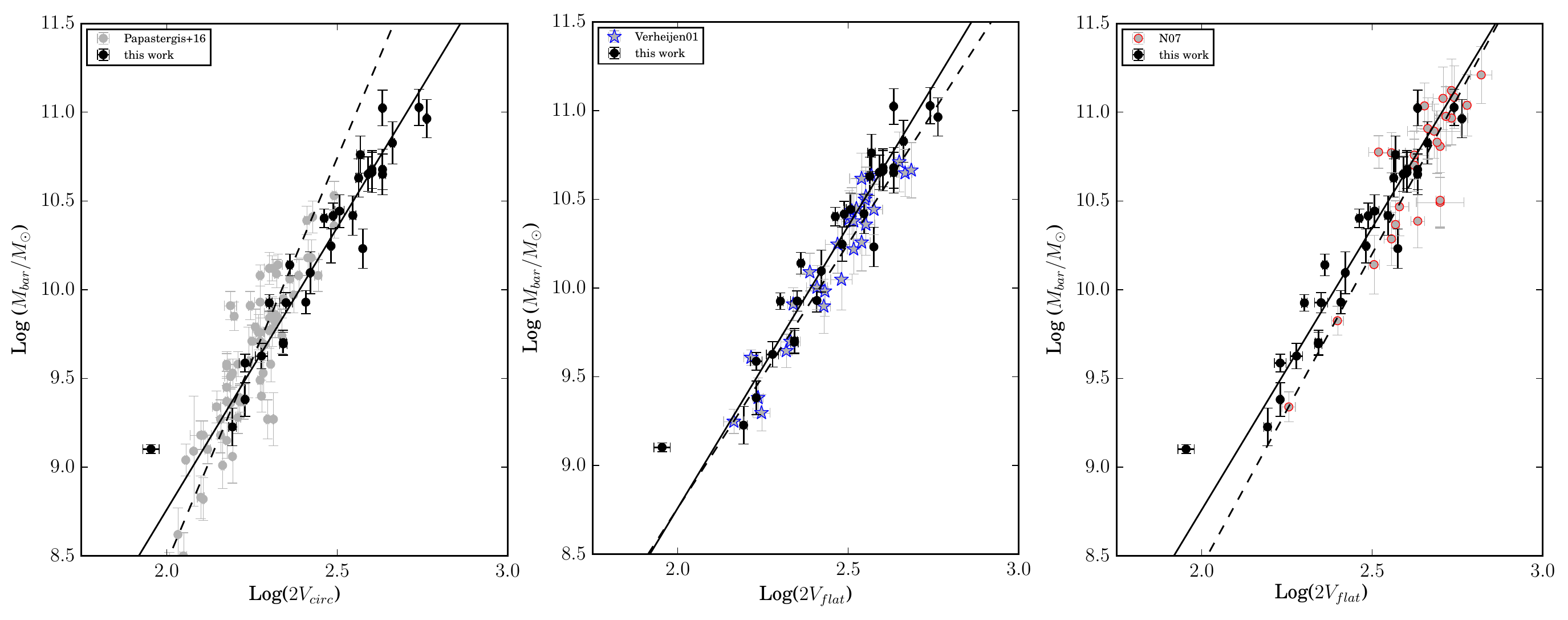}
\caption{ The comparison between our BTFr sample and previous studies:
  left panel from \citet{manolistfr}; middle panel from \citet{v01}
  and right panel from \citet{N07}.  With solid lines the fits for our
  sample are shown and with dashed lines the fits for previous studies
  are shown.}
\label{fig_iman}
\end{center}
\end{figure*} 


The biggest challenge in comparing our measurements of the statistical
properties of the BTFr with other studies is posed by the different
methods used to derive the main galaxy properties such as their
baryonic mass and rotational velocity \citep{bradford16}.  For
instance, the galaxy sample, the mass range of galaxies, the applied
corrections and the choice of the fitting method contribute
significantly to the measurement uncertainties.  Moreover, it is
critical in the comparison that the rotational velocities are
similarly defined.  Therefore, it is important to note that we
consider our $M_{bar,fin}-2V_{flat}$ relation for these comparisons.
However, it is not always the case that previous studies of the BTFr
are based on $2V_{flat}$ as a rotational velocity measure. In the
literature, it is more common that the global H{\sc i} profile widths
are used to estimate the circular velocity and, therefore, we further
refer to the rotational velocity in general as $V_{circ}$.


Our results are in a good agreement with those by \citet{lelli16} and
\citet{mcgaugh12} who both use $V_{circ}=V_{flat}$.  We obtain the
 observed vertical scatter $\sigma=0.16$ dex to be similar with the one
reported by \citet{lelli16} and our observed tightness $\sigma_{\perp}=0.052$ dex
is consistent with the total tightness $\sigma_{\perp}=0.06$ dex found
by \citet{mcgaugh12}. However, we find a shallower slope than the above mentioned studies. 
The slope of the BTFr reported by both \citet{lelli16} and
\citet{mcgaugh12} is measured to be $a\approx4$, while we find the
slope of the BTFr to be $a\approx3$. This can
be understood as they applied a higher stellar mass--to--light ratio
($\Upsilon_{\star}^{3.6}=0.5$ for the disk) which reduces the relative
contribution of the gas to M$_{bar}$ such that their BTFr approaches
our $L_{[3.6]}-2V_{flat}$ relation. They also do not include the presence of
the molecular gas in intermediate-- and high--mass spirals in their study.
Our slope is more consistent
with the result by \citet{zaritsky14}, who find the slope of the BTFr
to be in the range from $a=3.3$ to $a=3.5$, however they used the
corrected width of the global H{\sc i} profile $V_{circ}=W_{50}^{i}/2$.

\begin{table*}
\begin{tabular}{lcccc}
\hline
    & This work& Verheijen01& Noordermeer+07&Papastergis+16\\
\hline                
slope                 &2.99&2.98&3.51&4.58\\
zero point          &2.88 &2.82&1.4&-0.66\\
$\sigma$           &0.16&0.12&0.20&0.23\\
$\sigma_{\perp}$ &0.052&0.045&0.058&0.053\\
\hline
\end{tabular}
\caption{
Slope, zero point, scatter and tightness of the BTFrs
from different studies
Column(1): name of the parameter;
Column (2): slope, zero point, scatter and tightness from this work;
Column (3): slope, zero point, scatter and tightness from \citet{v01};
Column (4): slope, zero point, scatter and tightness from \citet{N07};
Column (5): slope, zero point, scatter and tightness from \citet{manolistfr}.}
\label{tbl_studies}
\end{table*}  

The low mass end of the BTFr is not well populated by our
sample. Moreover, the low--mass galaxies tend to have rotation curves
that are still rising at their outermost measured point and,
therefore, can not be considered for the $M_{bar,fin}-2V_{flat}$
relation.  Furthermore, the only low--mass galaxy in our sample (NGC
2366) appears to be very gas rich (see Figure \ref{fig_finbtfr}).  If we remove this galaxy 
from the fit, the statistical properties of our BTFr change slightly, as the
observed tightness changes from $\sigma_{\perp}=0.052$ to
$\sigma_{\perp}=0.048$ dex. However, there is nothing unusual about NGC 2366 that
would allow us to remove it from the BTFr. Moreover, it clearly demonstrates that inclusion of 
the gaseous component in the TFr increases its scatter, 
as NGC 2366 is not an outlier on the luminosity--based TFr of our sample.

\begin{figure*}
\begin{center}
\includegraphics[scale=0.55]{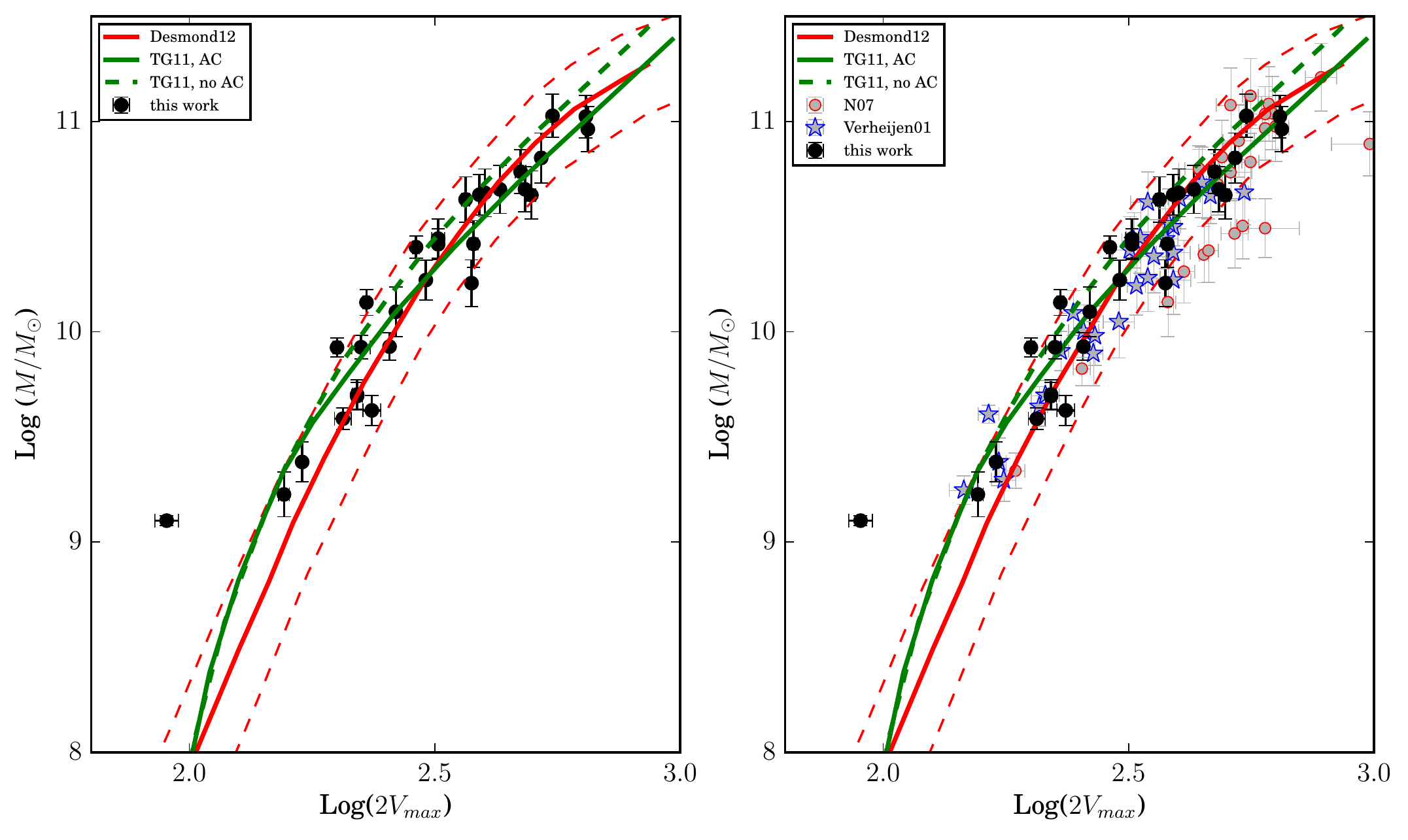}
\caption{Left panel: The comparison between our BTFr with the SAMs
  from \citet{TG11} (with adiabatic contraction (solid green line) and
  without (dashed green line)) and with the semi--analytic model from
  \citet{desmond12} (red line). With dashed red lines the 2$\sigma$
  intrinsic scatter is indicated. Right panel: the same comparisons
  but of the combined sample with \citet{N07} (red circles) and
  \citet{v01} (blue stars).}
\label{fig_tfsams}
\end{center}
\end{figure*}

For a more detailed comparison with previous studies, we present the
comparison analysis of the statistical properties of our BTFrs with
the BTFr from \citet{manolistfr, v01} and \citet{N07}.  The sample
from \citet{manolistfr} is a sample of heavily gas--dominated
($M_{gas}/M_{\star}\geqslant 2.7$) galaxies that populate the
low--mass end of the BTFr ($M_{\star}$ in a range from
$3\times 10^{8}$ to $ 3 \times 10^{10}$M$_{\odot}$), where $V_{circ}$
was measured as $V_{circ}=W_{50}^{i}/2$.  From this sample we adopt
only those 68 galaxies that have a large negative kurtosis ($h4<-1.2$)
of their global {H\sc i} profile, indicating that this profile is
double--peaked. Therefore, in these galaxies $W_{50}^{i}/2$ is
most likely a good approximation of $V_{flat}$.  The sample from
\citet{v01} is the Ursa Major sample of intermediate mass galaxies
($M_{\star}$ in the range from $1.6\times 10^{9}$ to
$4.0\times 10^{10}$M$_{\odot}$) where $V_{circ}=V_{flat}$ was measured
in the same manner as in our study, and
$M_{\star}=0.4L_{\star,K_{s}}$. The \citet{N07} sample is a sample of
 higher mass galaxies ($M_{\star}$ in the range from
$2.0 \times10^{9}$ to $2.0\times10^{11}$M$_{\odot}$) where
$V_{circ}=V_{flat}$ and $M_{\star}=0.4L_{\star,K_{s}}$ as well. It is
critical for the comparison of the BTFr studies that the fitting
algorithm is defined similarly \citep{bradford16}. Therefore, we apply
our fitting routine to the above--mentioned samples from
\citet{manolistfr, v01} and \citet{N07}, in order to derive the
statistical properties of the BTFrs in the same way as we did in our
study.  In Figure \ref{fig_iman} we present the comparisons between
our sample and those three from the literature.  The low--mass end and
the high--mass end samples show different results in comparison to our
BTFr.  For instance, the slope of the high--mass end sample
\citep{N07} is equal to $a=3.51$, while the slope of the low--mass
sample \citep{manolistfr} has the value of $a=4.58$.  Meanwhile, the
statistical properties of the \citet{v01} Ursa Major BTFr and our BTFr
are in excellent agreement, as the data for both these samples have
been treated in the same way (see Figure \ref{fig_iman} and Table
\ref{tbl_studies}).

The other important issue to keep in mind is that the baryonic masses
of galaxies were measured in different ways in the various
studies. This may also contribute to the differences we find while
performing these comparisons. For example, \citet{manolistfr} use the
average value of the stellar mass as derived with five different
methods for each galaxy in the sample \footnote[1]{1. using
  SED--fitting of the SDSS bands; 2. using SED--fitting of the SDSS +
  GALEX bands; 3. using $g-i$ colour; 4. using constant
  $\Upsilon_{\star}=0.45$ in 3.4 $\mu$m band; 5. using constant
  $\Upsilon_{\star}=0.6$ in K--band.} and $M_{gas}=1.3 \times
M_{HI}$.
\citet{zaritsky14} adopt a stellar mass as measured from the ratio of
the 3.6 and 4.5 $\mu$m fluxes.  \citet{lelli16} and \citet{mcgaugh12}
adopt a single mass--to--light ratio for all galaxies equal to
$\Upsilon_{\star}^{3.6}=0.5$ for the disk. The results for our BTFr
with the adopted stellar mass--to--light ratio equal to
$\Upsilon_{\star}^{3.6}=0.5$ are presented in Section \ref{btfrscomp},
and we recall that this choice of $\Upsilon_{\star}^{3.6}=0.5$ for our
sample also results in a shallower slope of $a=3.15\pm0.21$ for the
BTFr.

\subsection{Semi--analytical models}

In the DM--only simulations within the $\Lambda$CDM cosmological model
framework the mass of the DM halo relates to its rotational velocity
as $M_{h} \propto V_{h, max}^{3}$. Therefore, the BTFr just follows
from this relation as $M_{bar} \propto V_{rot}^{3}$ \citep{klypin11}.
A further study of the BTFr in the $\Lambda$CDM context is done by
using semi--analytical models (SAMs) of galaxy formation which assign
observationally motivated masses of stars and gas to the host dark
matter halo, and $V_{circ}$ is usually calculated by computing the
rotation curve, which takes into account the contribution of the
baryons to the total gravitational potential from which the rotation
curve is calculated \citep{moster10,TG11}. Consequently, the value of
$V_{circ}$ is either determined at a particular radius or is assumed
to be the rotational velocity of the peak of the simulated rotation
curve $V_{circ}=V_{max}$.
 

\begin{figure}
\begin{center}
\includegraphics[scale=0.6]{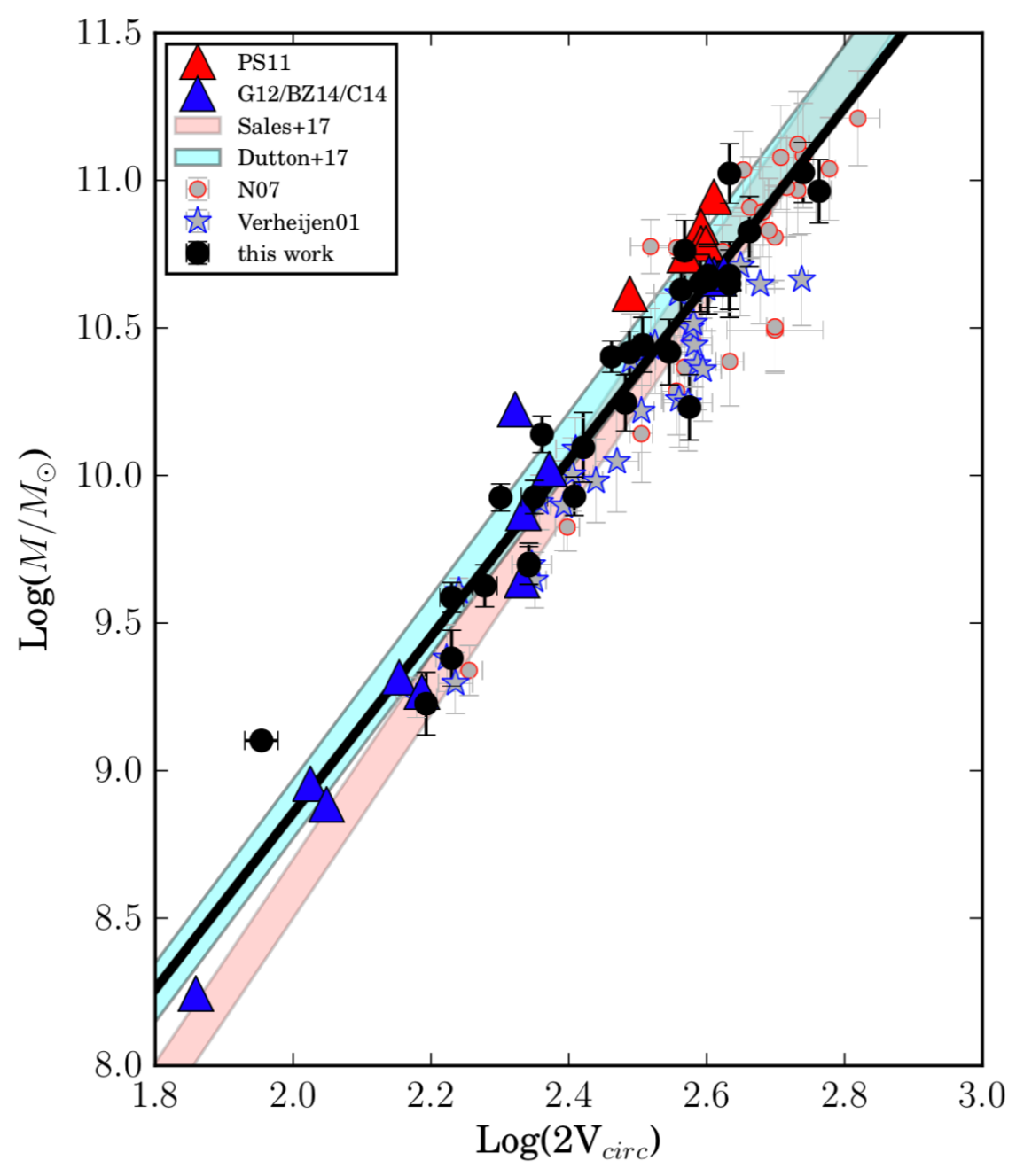}
\caption{ The comparison of our BTFr combined with \citet{N07} (red
  circles) and \citet{v01} (blue stars) samples, with the galaxies
  produced by the hydrodynamical simulations from \citet{piontek11}
  (red triangles), and by \citet{governato12, brooks14} and
  \citet{christ14} (blue triangles). The best-fit
  BTFr with the measured scatter from the APOSTLE/EAGLE simulations is
  shown with red shaded area\citep{sales17}. The NIHAO simulations best-fit BTFr with the measured scatter is shown with
  the cyan area \citep{dutton17}.}
\label{fig_hydro}
\end{center}
\end{figure} 

We compare our BTFr with those from two SAMs from \citet{TG11} and
from \citet{desmond12}. From \citet{TG11} we consider two models: one
in which DM halos experience adiabatic contraction due to the infall
of the baryons to the halo centres, and one without adiabatic
contraction. The rotation velocity in \citet{TG11} is calculated at a
fixed radius of 10 kpc.  The semi--analytical model of
\citet{desmond12} uses the $V_{circ}=V_{max}$ estimation of the
rotational velocity.  Moreover, the \citet{desmond12} model also
calculates the intrinsic scatter of the BTFr, expected in a
$\Lambda$CDM cosmology. This scatter is caused by various mechanisms,
such as scatter in the concentration parameter of the DM halos,
scatter in the halo spin parameter and scatter in the baryon fraction
of the halo.  Figure \ref{fig_tfsams} presents the comparison of our
sample BTFr (left panel) and of the combined sample with those from
\citet{v01} and \citet{N07} (right panel) with these two models.  Note
that for the observational samples we use $V_{circ}=V_{max}$ for a
fair comparison.  The slight curvature in the BTFr that is present in
both models is a general prediction of SAMs in the $\Lambda$CDM
cosmological model. Both models can reproduce our sample relatively
well (Figure \ref{fig_tfsams}) with the one clear outlier NGC 2366.
Even though the \citet{desmond12} model is systematically offset from
the observed data points of all three samples, it can reproduce the
relations within the $2\sigma$ uncertainty accounting for the
intrinsic scatter of the BTFr, as indicated with the dashed red lines
in Figure \ref{fig_tfsams}. It is important to note that, even though
SAMs can reproduce our BTFr with $2\sigma$ uncertainty, we do not 
find the presence of curvature in our observed BTFr.

\subsection{Hydrodynamical simulations}

In Figure \ref{fig_hydro} we compare our BTFr with individual galaxies
produced by hydrodynamical simulations in the context of a
$\Lambda$CDM cosmological model and with the best-fit BTFr with the
measured scatter from the APOSTLE/EAGLE (red shade,
\citealp{sales17}) and the NIHAO (cyan shade, \citealp{dutton17}) hydrodynamical simulations.  
While both sets of the simulations are successful in reproducing the high--mass end
of our BTFr, the NIHAO works better on the smaller scales. 
From individual galaxies for the high--mass end of the BTFr we consider eight
galaxies from \citet{piontek11} and for the intermediate and low--mass
end we consider twelve galaxies, produced in a set of hydrodynamical
simulations by \citet{governato12, brooks14} and \citet{christ14}. In
Figure \ref{fig_hydro} it is shown that these sets of hydrodynamical
simulations are successful at reproducing the observed BTFr with the
exception of the \citet{piontek11} sample of high mass galaxies, which
lies systematically above the observed BTFr. The rotational velocities of galaxies in these
simulations are measured from the width of the global H{\sc i}
profiles and therefore, for z fair comparison, we also use $W_{50}$ as a
rotational velocity measure for the observational samples.


\begin{figure}
\begin{center}
\includegraphics[scale=0.6]{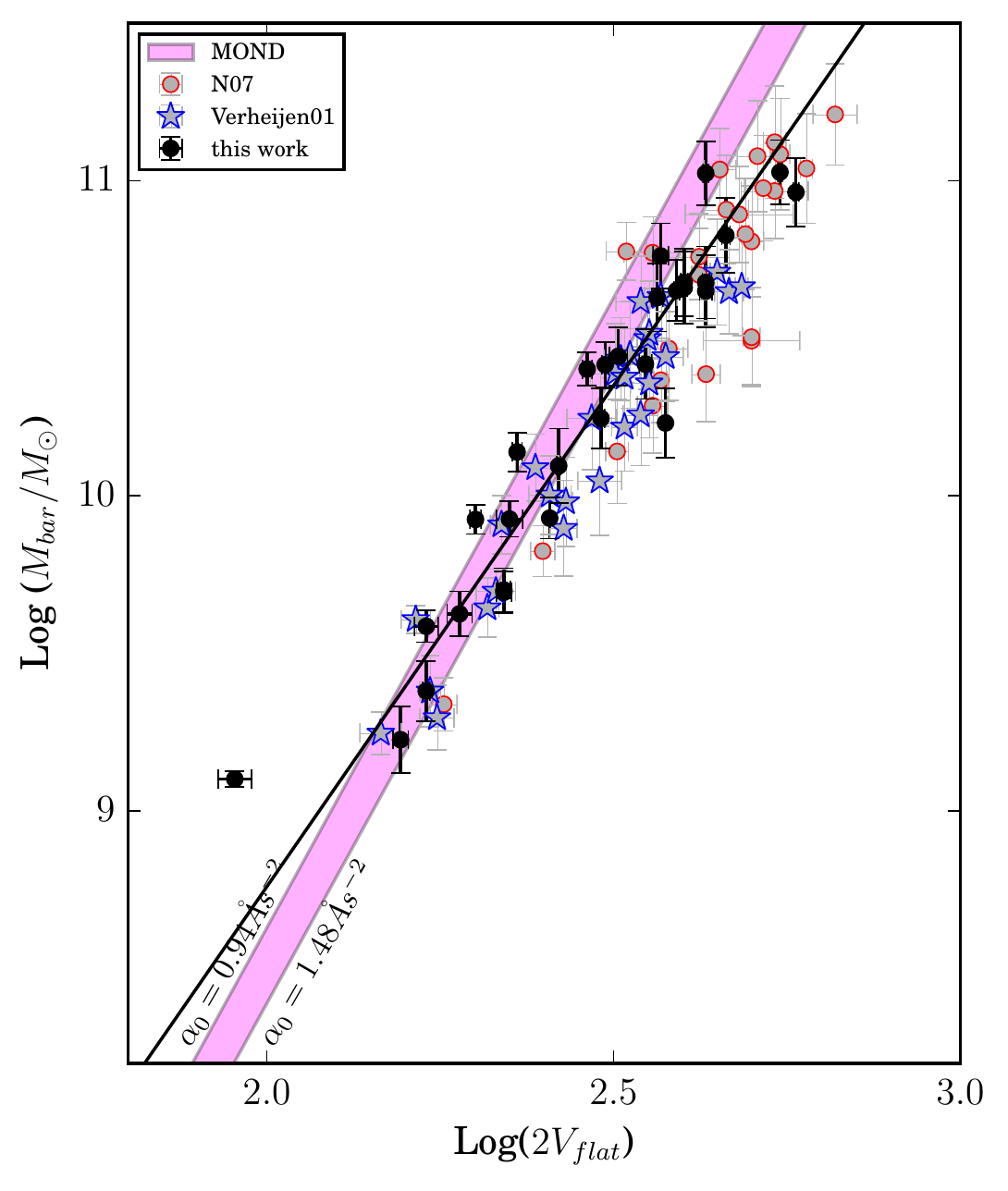}
\caption{ Our BTFr combined with the samples from \citet{N07} and
  \citet{v01} and in comparison with the BTFr predicted by MOND.}
\label{fig_tfmond}
\end{center}
\end{figure}   

\subsection{MOND} 

Modified newtonian dynamics (MOND) is an alternative to the
$\Lambda$CDM model of galaxy formation and evolution, which does not
require the presence of dark matter, but where gravitational forces
are entirely defined by the amount and distribution of the baryonic
matter \citep{milgrom88}.  Therefore, it predicts that the BTFr with
$V_{circ}=V_{flat}$ can be described with a single power--law with
zero intrinsic scatter, and the relation has a slope of exactly 4 when
it is based on $V_{flat}$ as a rotational velocity measure, while the
normalisation of the relation depends only on the acceleration
parameter $\alpha_{0}$ \citep{mcgaugh12}.

The advantage of our sample is that it is based on $V_{flat}$ as a
rotational velocity measure and therefore allows us to directly
compare our results with the predictions from MOND. 
While the slope and zero intrinsic scatter of the BTFr from the MOND
prediction are fixed, the normalisation of the relation can vary due
to the uncertainty in the observational determination of the
acceleration parameter $\alpha_{0}$ \citep{begeman91}. In Figure
\ref{fig_tfmond} we consider two values for the acceleration parameter
$\alpha_{0}=0.94\mathring{A}s^{-2}$ and
$\alpha_{0}=1.48\mathring{A}s^{-2}$ which are consistent with
\citet{begeman91}. 
Since we deal mainly with bright galaxies, where the effect of 
MOND is less obvious than for the dwarfs, we should 
note that we do not probe deeply into the MOND regime at the 
faint end of the BTFr. However, from Figure \ref{fig_tfmond} it is clear that,
while the MOND normalisation works well for the intermediate mass
galaxies, it fails to reproduce the high--mass end of the BTFr
(represented by our, \citet{v01} and \citet{N07} samples) by
introducing a slope that is too steep.  In conclusion, neither the
non--zero intrinsic scatter nor the slope of our BTFr are consistent
with the relation predicted by MOND.


\section{Summary and Conclusions}\label{concl}

In this paper we perform a detailed study of the Baryonic
Tully--Fisher relation based on different stellar mass estimates and
taking advantage of a sample with accurate primary distance
measurements, homogeneously analysed photometry, and spatially
resolved H{\sc i} kinematics. The aim of our study is to investigate
how the various stellar mass estimation methods affect the statistical
properties of the BTFr. For this, we estimate the stellar masses of
our sample galaxies following four different prescriptions. First, we
measured stellar masses by performing a full SED--fitting using 18
photometric bands (from FUV to far infrared).  Second, we adopt a
median value of the dynamical mass--to--light ratio from the DiskMass
survey $\Upsilon_{\star}^{Dyn, K}=0.3$. Then, we calculate the stellar
mass--to--light ratio for the 3.6 $\mu$m band as a function of
[3.6]--[4.5] colour and, finally, we adopt the single mass--to--light
ratio equal to $\Upsilon_{\star}^{[3.6]}=0.5$ from \citet{lelli16},
which is motivated by an empirical minimisation of the vertical
scatter in the BTFr.

Using each stellar mass estimate, we construct the Baryonic
Tully--Fisher relations. Each of the relations is based on three
different velocity measures: $W_{50}$ from the global H{\sc i}
profile, and $V_{max}$ and $V_{flat}$ from the rotation curve.  For
each of the relations we measure the slope, vertical scatter and
tightness. We find that the tightest BTFrs with the smallest vertical
scatter are those based on larger mass--to--light ratios (method 3 and
method 4) and based on $W_{50}$ as a rotational velocity
measure. However, none of the relations demonstrates as small a
vertical and perpendicular scatter as the 3.6 $\mu$m luminosity--based
TFr. This allows us to conclude that mostly the gas component is
responsible for the increase of the scatter (vertical and
perpendicular) in the BTFr. Hence, increasing the mass--to--light
ratio reduces the vertical and perpendicular scatter of the BTFr, as
it makes the gas contribution negligible.

We consider in detail our BTFr of choice, which is based on $V_{flat}$
and a stellar mass computed with a single mass--to--light ratio
$\Upsilon_{\star}^{[3.6]}=0.35$. This choice of the mass--to--light
ratio is motivated by adopting a value intermediate between the high
values of methods 3 and 4 and low values of methods 1 and 2.  We
measure the slope, vertical scatter and tightness of our BTFr. We find
the slope equal to 2.99$\pm$0.28, which is shallower in comparison
with previous studies \citep{lelli16, mcgaugh12}, and the vertical
scatter $\sigma=0.16$ dex, which is consistent with the previous study
by \citet{lelli16}.  We find the observed tightness
$\sigma_{\perp}=0.052$ dex to be smaller than the ones found by
\citet{manolistfr} and \citet{mcgaugh12}. This observed perpendicular
scatter is larger than the perpendicular scatter of the 3.6 $\mu$m
luminosity--based TFr \citepalias{pon17}. However, the estimated intrinsic perpendicular
scatter is shown to be similar $\sigma_{\perp,int} \sim 0.026 \pm 0.13$ dex.

Furthermore, we compare the results of our BTFr with various
theoretical predictions from $\Lambda$CDM theories of galaxy formation
and MOND. We find that semi--analytic models \citep{TG11, desmond12} and different sets of 
cosmological hydrodynamical simulations \citep{dutton17, sales17} represent our relation reasonably 
well within the allowed 2$\sigma$
uncertainty. Moreover, individual galaxies of different masses from various
hydrodynamical simulations \citep{governato12,brooks14,christ14,
  piontek11} also tend to follow our observed BTFr. However, the
predictions from MOND fail to reproduce the observed BTFr at the
high--mass end, where the observed galaxies of high masses tend to lie
below the MOND predicted relation since the slope predicted by MOND is
steeper. Our observed BTFr is shown not to be consistent with a zero
intrinsic scatter at the $2\sigma$ level.

In conclusion, it is important to point out that there is no unique
method to estimate the stellar masses of spiral galaxies. Various
methods lead to estimates which may differ significantly from each
other.  Therefore, the statistical properties of the Baryonic
Tully--Fisher relation remain uncertain as different stellar
mass--to--light ratios lead to different estimates of the slope,
scatter and tightness of the relation.




\section*{acknowledgements}

AP is grateful to Jayaram Chengalur, Filippo Fraternali and Federico
Lelli for useful comments and suggestions. AB acknowledges financial
support from the CNES (Centre National d'Etudes Spatiales -- France). EP
is supported by a NOVA postdoctoral fellowship of the Netherlands
Research School for Astronomy (NOVA). MV acknowledges the Netherlands 
Foundation for Scientific Research support through VICI grant 016.130.338.
We acknowledge the Leids Kerkhoven--Bosscha Fonds (LKBF) for travel support.
We acknowledge financial support from the DAGAL network
from the People Programme (Marie Curie Actions) of the European Union's
Seventh Framework Programme FP7/2007-2013/ under REA grant agreement
number PITNGA-2011-289313.
We thank Elisabete da Cunha for making the MAGPHYS SED--fitting code publicly available.

\bibliographystyle{mnras.bst} 
\bibliography{Chap4} 

\appendix
\section{Uncertainties in the SED--fitting}
\label{appendix:sed1}
\begin{figure}
\begin{center}
\includegraphics[scale=0.6]{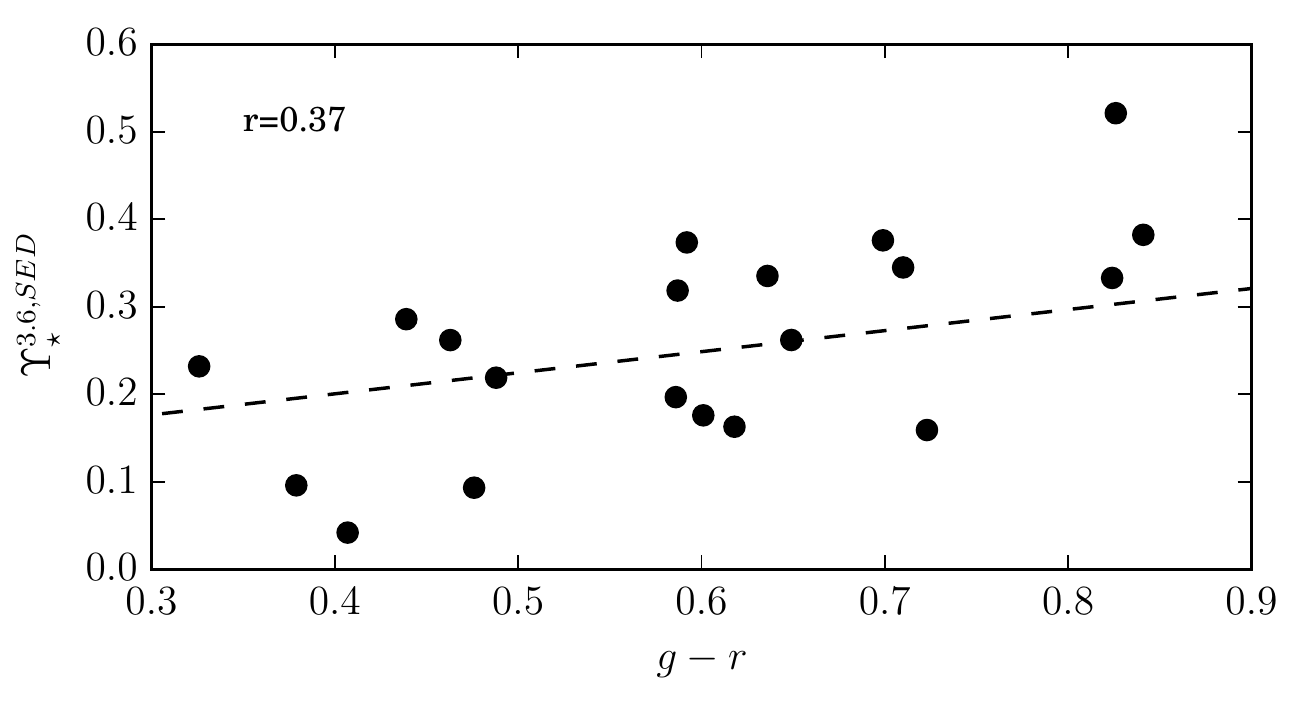}
\caption{Correlation between $\Upsilon_{\star}^{SED, [3.6]}$ and $g-r$ colour. $r$ is Pearson's correlation coefficient.
\label{fig_sdssml}}
\end{center}
\end{figure}

\begin{figure}
\begin{center}
\includegraphics[scale=0.6]{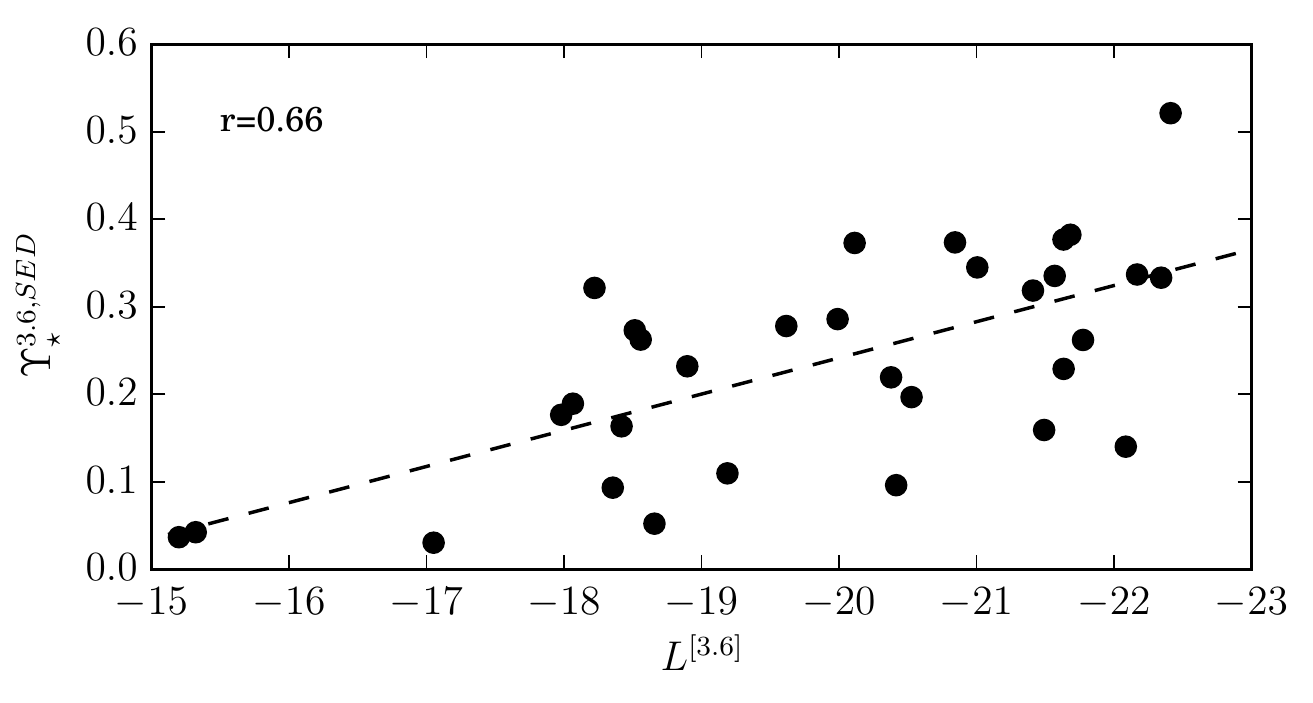}
\caption{Correlation between $\Upsilon_{\star}^{SED, [3.6]}$ and total luminosity of a galaxy in  3.6 $\mu$m band. $r$ is Pearson's correlation coefficient.
\label{fig_mlm36}}
\end{center}
\end{figure}

It is clearly demonstrated in Figures \ref{fig_seddm} and \ref{fig_sedcor} that 3.6 $\mu$m stellar mass--to--light ratios, measured from the SED--fitting span a wide range, while stellar population synthesis models predict this value to be more or less constant. The major concern regarding the 
SED--fitting results is that for several objects mass--to--light ratios are very small, for example $\Upsilon_{\star}^{SED, [3.6]} = 0.03 $ for IC2574 or  $\Upsilon_{\star}^{SED, [3.6]} = 0.04 $ for NGC 2366.  If these were real, the galaxies would produce 10 times more light for the same mass, and this should be seen strongly in the colours, unless most of it is obscured. Figures \ref{fig_sdssml} and \ref{fig_mlm36} show the correlation between $\Upsilon_{\star}^{SED, [3.6]}$ and $g-r$ colour and total luminosity of a galaxy in  the 3.6 $\mu$m band ($L_{[3.6]}$), respectively. While the correlation between $\Upsilon_{\star}^{SED, [3.6]}$ and $g-r$ colour is absent with Pearson's coefficient $r$ equal to 0.37, $\Upsilon_{\star}^{SED, [3.6]}$ correlates well with the $L_{[3.6]}$ ($r=0.66$). This correlation with the band which is closely related to the stellar mass illustrates that mostly low mass dwarf galaxies have low mass--to--light ratios. The possible reason is the fact that there might be contamination of the 3.6 $\mu$m flux due to non-stellar emission and by the leakage of the PAH bands. However, the low mass--to--light ratios of these galaxies do not directly result into an increased scatter of the BTFr, as many of them have rising rotation curves and were not included in the statistical analysis. 

\begin{figure}
\begin{center}
\includegraphics[scale=0.55]{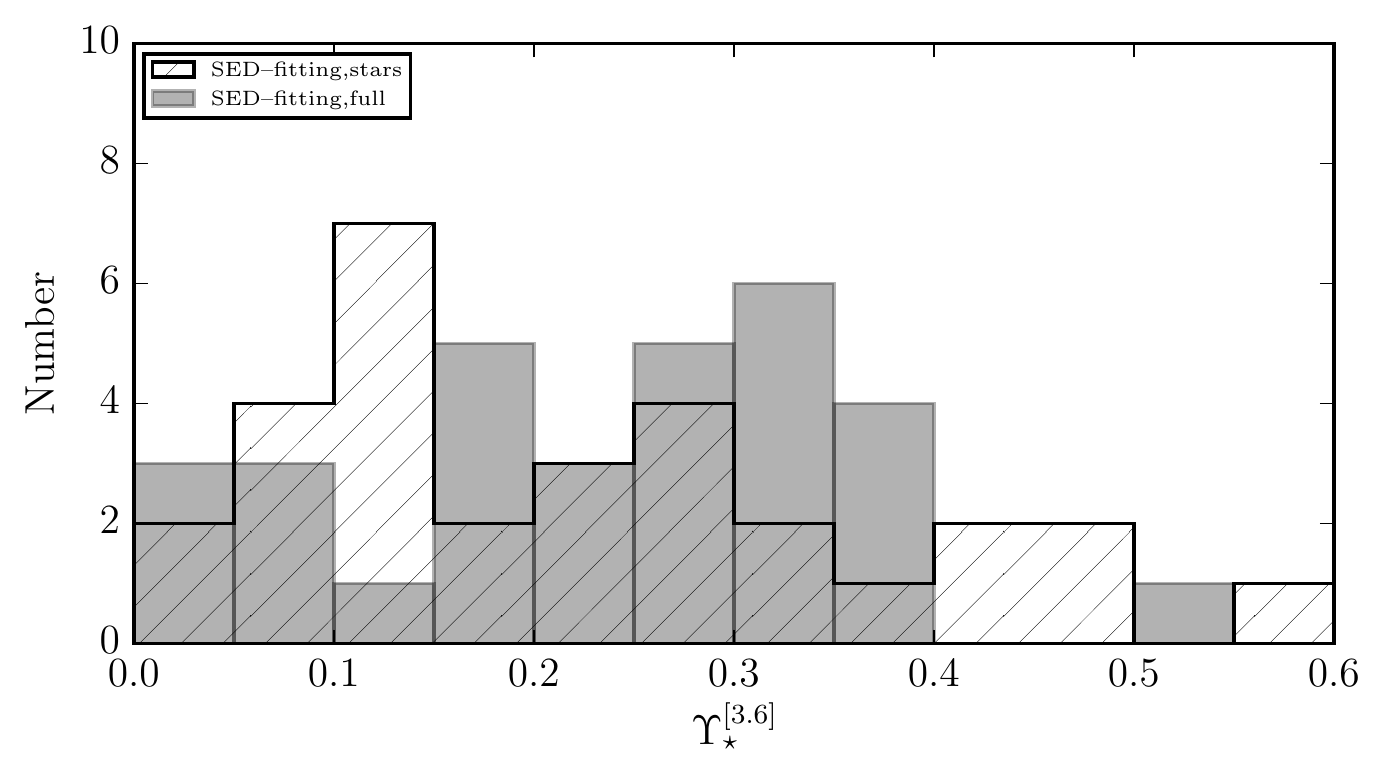}
\caption{A comparison between $\Upsilon_{\star}^{SED, [3.6]}$ obtained using information from all 18 bands (grey area) and using only stellar bands $griz$ \& $JH$ (hatched area).
\label{fig_fullstars}}
\end{center}
\end{figure}

\begin{figure}
\begin{center}
\includegraphics[scale=0.85]{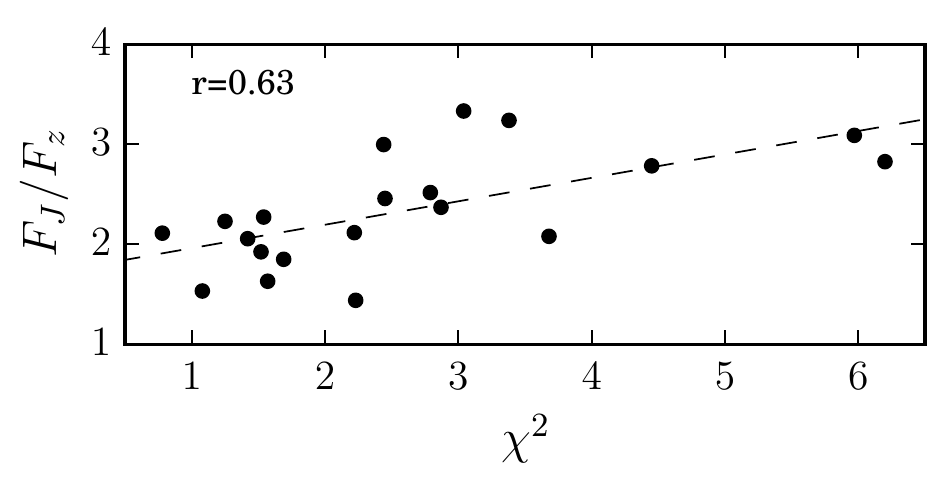}
\caption{Correlation between the difference of fluxes in $z$-- and $J$-- bands and the minimum $\chi^{2}$ for the best--fit SED--fitting model. 
$r$ is Pearson's correlation coefficient.
\label{fig_zjchi2}}
\end{center}
\end{figure}

Another concern is that the best errors we can apply to the stellar mass--to--light resulting from the SED--fitting are motivated by the study of 
\citet{roediger15} who found a very low error of 0.1 dex using a set of simulated galaxies. However, \citet{roediger15} only used $griz$ and $H$-- band for their study, basically only bands, dominated by the stars. Unlike us, they did not use any dust emission. The inclusion of the dust emission is an important constraint as it regulates the amount of extinction in the bands dominated by stellar light. Figure \ref{fig_fullstars} shows a comparison between $\Upsilon_{\star}^{SED, [3.6]}$ obtained using information from all 18 bands and using only the stellar $griz$ \& $JHK$ bands, similar to \citet{roediger15}. It is clearly seen from Figure \ref{fig_fullstars} that use of only stellar bands still produces a wide spread of $\Upsilon_{\star}^{SED, [3.6]}$. Moreover, the amount of galaxies with very low mass--to--light ratios increases. Figure \ref{fig_zjchi2} shows the correlation ($r=0.63$) between the difference of fluxes in the $z$-- and $J$-- bands and the minimum $\chi^{2}$ for the best--fit SED--fitting model for each galaxy. The correlation between other bands flux ratios and $\chi^{2}$ is absent. 

\section{SED best--fit models for our sample galaxies}\label{appendix:sed}

In Figures B1-- B32 we present the best--fit models performed with
MAGPHYS (in black)  over the observed spectral energy distribution of
our sample galaxies (see Figure \ref{fig_sed} for an example). The blue curve shows the unattenuated stellar
population spectrum.  The bottom plot shows the residuals for each
measurement ($(L_{obs}-L_{mod})/L_{obs}$). The full Appendix B data can be found \href{https://www.dropbox.com/s/h389tfjl8dx0rpd/AppendixB.pdf?dl=0}{here}.


\end{document}